\documentclass[fleqn,usenatbib]{mnras}

\usepackage{framed}
\usepackage{graphicx,epsfig,psfig,multicol}
\usepackage[dvipsnames]{xcolor}
\usepackage[]{inputenc,amssymb}
\usepackage{amsmath}
\usepackage{color}
\usepackage{epsfig}
\usepackage{epstopdf}
\usepackage{textcomp}
\usepackage{xcolor,cancel}
\usepackage{comment}
\usepackage{subcaption}

\usepackage[export]{adjustbox}
\usepackage{adjustbox}
\usepackage{placeins}

\usepackage{float}
\usepackage{morefloats}

\usepackage{lipsum}
\usepackage{mwe}
\usepackage{arydshln} 

\definecolor{webgreen}{rgb}{0,.5,0}
\definecolor{webbrown}{rgb}{.6,0,0}

\usepackage[toc,page]{appendix}

\def \beq{\begin{equation}}
\def \eeq{\end{equation}}
\def \bea{\begin{eqnarray}}
\def \eea{\end{eqnarray}}

\newcommand{\rev}[1]{\textcolor{black}{#1}}

\usepackage[T1]{fontenc}

\DeclareRobustCommand{\VAN}[3]{#2}
\let\VANthebibliography\thebibliography
\def\thebibliography{\DeclareRobustCommand{\VAN}[3]{##3}\VANthebibliography}

\title[Loss Cone Shielding]{Loss Cone Shielding}

\author[O. Teboul et al.]{
Odelia Teboul$^1$\thanks{E-mail: odelia.teboul1@mail.huji.ac.il},
Nicholas C. Stone$^1$, Jeremiah P. Ostriker$^2$
\\
 $^1$Racah Institute of Physics, The Hebrew University, 91904, Jerusalem, Israel
\\
 $^2$Department of Astronomy, Columbia University, 550 West 120th St, New York, NY 10027, USA}

\date{Accepted 2023 October 23. Received 2023 October 03; in original form 2022 November 17}

\pubyear{2022}

\begin{document}
\twocolumn
\label{firstpage}
\pagerange{\pageref{firstpage}--\pageref{lastpage}}
\maketitle

\begin{abstract}

A star wandering close enough to a massive black hole (MBH) can be ripped apart by the tidal forces of the black hole. The advent of wide-field surveys at many wavelengths has quickly increased the number of tidal disruption events (TDEs) observed, and has revealed that i) observed TDE rates are lower than theoretical predictions and ii) E+A galaxies are significantly overrepresented. This overrepresentation further worsens the tension between observed and theoretically predicted TDEs for non-E+A galaxies. Classical loss cone theory focuses on the cumulative effect of many weak scatterings. However, a strong scattering can remove a star from the distribution before it can get tidally disrupted. Most stars undergoing TDEs come from within the radius of influence, the densest environments of the universe. In such environments, close encounters rare elsewhere become non-negligible. We revise the standard loss cone theory to take into account classical two-body interactions as well as strong scattering, collisions, tidal captures, and study under which conditions close encounters can shield the loss cone.  We i) analytically derive the impact of strong scattering and other close encounters, ii) compute time-dependent loss cone dynamics including both weak and strong encounters, and iii) derive analytical solutions to the Fokker-Planck equation with strong scattering.  We find that i) TDE rates can be reduced to up to an order of magnitude and ii) strong shielding preferentially reduces deeply plunging stars.  We also show that stellar overdensities, one possible explanation for the E+A preference, can fail to increase TDE rates when taking into account strong scattering.

\end{abstract}

\section{Introduction}
\label{sec:intro}
A star wandering too close to a massive black hole (MBH) can be ripped apart by the MBH's tidal forces. Roughly half of the gaseous debris from the disrupted star falls back onto the MBH, eventually circularizing into an accretion disk and powering a luminous flare \citep{Rees88, EvansKochanek89}.

Tidal disruption events (TDEs) have been theoretically predicted since the 1970s (e.g., \citealt{Hills75}) but were not observed until the advent of all-sky surveys \citep[beginning with the ROSAT X-ray all-sky survey;][]{Bade, Komossa}. The sample of observed TDEs has rapidly grown in the last decades and now spans all wavelengths, from the radio to gamma-rays.  The last decade has seen a high rate of TDE discovery in wide-field optical surveys such as SDSS \citep{vanVelzen11}, Pan-STARRS \citep{Gezari, Chornock,Nicholl}, PTF \citep{Arcavi+14}, iPTF \citep{Blagorodnova, Hung}, ASASSN \citep{Holoien16, Hinkle}, and ZTF \citep{vanVelzen+19, Hammerstein22}.  X-ray surveys are also returning to prominence as a driver of TDE discovery; at present, the SRG/eROSITA all-sky survey \citep{Sazonov} seems to be finding $\sim 20$ TDEs per year, comparable to the optical TDE discovery rate of ZTF. As time domain capabilities continue to expand, the observed sample of TDEs is expected to grow by orders of magnitude, with hundreds to thousands of new TDEs expected to be found in optical wavelengths by the LSST/Vera Rubin Observatory \citep{Ivezi, BricmanGomboc20}, 
and in the near-ultraviolet by ULTRASAT \citep{Ultrasat}. 

The opening of this new era for TDEs has allowed observers to empirically constrain the per-galaxy TDE rate to a range $\dot{N}_{\rm TDE} \sim 10^{-5} - 10^{-4}$ yr$^{-1}$~ gal$^{-1} $ \citep{Holoien16, vanVelzen18}, and revealed a surprising over-representation of these events in post-starburst (e.g. ``E+A'') galaxies \citep{Arcavi+14, French+16,French+17, Law, Graur, French20, Hammerstein21}. The latest estimations suggest that E + A galaxies are overrepresented by a factor of $ \approx 22 $ \citep{Hammerstein21}. 
This further reduces the observationally inferred TDE rate in non-E+A galaxies, which is already in tension with theoretical (but empirically calibrated) rate estimates of  $\dot{N} \sim 10^{-4} - 10^{-3}$ yr$^{-1}$~ gal$^{-1}$ (e.g., \citealt{WangMerritt04, StoneMetzger16}).

 TDE rates are traditionally estimated from classical loss cone theory, which focuses on the cumulative effect of many weak scatterings. These scatterings are modeled as local and uncorrelated, and lead to effective diffusion coefficients describing the drift and diffusion of stellar populations through phase space \cite[see][for reviews]{Merritt13, Stone+20}. Strong, or small impact parameter scatterings are generally neglected, as they are largely outnumbered by weak scatterings.  
 
 However, strong scatterings can have a qualitatively different effect than weak ones on stellar populations.  In particular, {\it a strong scattering can eject a star from the distribution} before it gets tidally disrupted.  Most stars undergoing TDEs come from within the radius of influence, {\it the densest environments in the Universe}. Such an environment can facilitate ejection in strong scatterings, but it also increases the rate of close encounters that are rare elsewhere. Such close encounters can be divided into star-star interactions, including collisions and tidal captures, and star-stellar mass black hole (sBH) interactions, such as $\mu$-TDEs and tidal captures. In this paper we will revise the standard loss cone theory to take into account not simply weak two-body scatterings, but also destructive processes such as strong scattering ejections, collisions,  and tidal captures, with the ultimate goal of determining under which conditions close encounters can {\it shield the loss cone}. 

Close encounters are highly dependent on both the stellar and sBH slopes. An important prediction of stellar dynamics is the redistribution of arbitrary initial distributions of stars around MBHs over time into quasi-stationary states (QSSs). In the presence of one population of stars, a Bahcall-Wolf density cusp with stellar density $\rho \propto r^{-\gamma} $, will develop, with $\gamma=7/4$ \citep{BahcallWolf76}. If the system is composed of both stars and heavier objects, the heavier objects are expected to segregate towards the centre of the galactic nucleus and settle on a steeper cusp $ \gamma = 7/4 - 2$  while the light objects will have a weaker cusp $\gamma \approx 1.3 -1.5 $ \citep{BW77, PretoAmaroSeoane10, Amaro, Broggi}. In the strong segregation limit, when the heavier objects are relatively rare, the heavy objects  have been predicted to settle to even steeper power-law slopes of $\gamma = 2 - 11/4  $ \citep{AlexanderHopman09}. \rev{Moreover for an  adiabatically growing MBH, the stellar slope may assume steeper values  of $\gamma \gtrsim 2$ \citep{Young}.}

We introduce the different kinds of close encounters that are relevant for our problem in \S \ref{sec:close}, and derive their analytical rates in \S \ref{sec:strong}.  In \S \ref{sec:loss}, we solve the Fokker-Planck equation in angular momentum space in the presence of close encounters. In \ref{sec:analytical} we present our analytical time-dependent solutions of the Fokker-Planck equation with strong scattering. In \S \ref{sec:results} we compute the impact of close encounters on TDE rates and discuss our findings for the E+A preference. In \S \ref{sec:conclusions}, we discuss the implications of our work and briefly summarize it.  

\section{Close Encounters}
\label{sec:close}

In a spherical galactic nucleus one can characterize the population of stars with an orbit-averaged distribution function \rev{ $f(\epsilon, J)$}, where $\epsilon$ is the orbital energy and $J$ the orbital angular momentum of stars.  A variety of processes can remove stars from the distribution $f$.  Most notably, a star of mass $M_\star$ and radius $R_\star$ will be destroyed by the central MBH (of mass $M_\bullet$) if it completes an orbit with galactocentric pericenter $R_{\rm p}$ less than the tidal radius,
\begin{equation}
    R_{\rm t} = R_\star \left( \frac{M_\bullet}{M_\star}\right)^{1/3}. \label{eq:rt}
\end{equation}
A finite $R_{\rm t}$ (i.e. the presence of a central MBH) carves a conical gap in velocity space (at fixed radius), or equivalently places a non-zero lower limit on the angular momentum that particle orbits may have \citep{FrankRees76, LightmanShapiro77, CohnKulsrud78}.  Particles with less angular momentum suffer destructive interactions at pericenter. 

However, before the typical star can get to an orbit with $R_{\rm p} \sim R_{\rm t}$, its pericenter must endure a long and potentially dangerous random walk through regions of high stellar and sBH density.  During this random walk, sufficiently close encounters with the stellar/sBH population can eliminate our test star through a variety of different processes.  Here we overview the relevant ones, roughly in order of increasing cross-section:
\begin{enumerate}
    \item Direct physical collisions: star-star (or star-sBH) collisions with a pairwise pericenter $p<R_\star$ will often destroy the test star.  At large distances $r$ from the MBH, where the velocity dispersion $\sigma \ll V_\star=\sqrt{2GM_\star/R_\star}$, the stellar surface escape speed, collisions with other stars can be coagulative rather than destructive\footnote{As we are focused on relatively small spatial scales, we neglect this possibility in this paper.}.
    
    \item $\mu$-TDEs: slightly more distant encounters with a sBH ($p < r_{\rm t}$, where $r_{\rm t}$ is the same as $R_{\rm t}$ from Eq. \ref{eq:rt} but with the sBH mass $m_{\rm bh}$ substituted into Eq. \ref{eq:rt} for $M$), can destroy the test star through the action of the sBH's tides, in a scaled down version of a standard TDE.  The observational signatures of these ``$\mu$-TDEs'' are poorly understood \citep{Perets+16}.  However, the different dynamics and greatly reduced Eddington limit of an sBH (compared to an MBH) means that the resulting transient is unlikely to be mistaken for a standard TDE, and a star experiencing this fate can be safely removed from the distribution function.
    \item Tidal captures: at slightly larger pericenter distances than the $\mu$-TDE radius, tides can excite mechanical oscillation modes in the test star.  If this process converts enough orbital energy into mechanical energy, then an unbound parabolic/hyperbolic two-body orbit will capture into a bound binary \citep{Fabian+75, PressTeukolsky77, LeeOstriker86}.  A tidal capture can only occur when the relative kinetic energy between the two objects is less than the binding energy of the test star.  The long-term evolution of a tidal capture system is not entirely clear \citep{Generozov+18}.  Some recent hydrodynamical simulations indicate that the captured star will be destroyed in a brief sequence of runaway partial (micro)-TDEs \citep{Kremer+22}, but even if this is not the case, a star that has become tightly bound to a sBH is unlikely to ever experience a TDE.  If a tight binary's center-of-mass angular momentum starts falling to levels approaching the loss cone, then direct merger (i.e. $\mu$-TDE) becomes the most probable outcome \citep{Bradnick+17}, followed by tidal separation via the Hills mechanism.  Star-star tidal captures are also possible, but again the most likely final result (if the binary is excited towards small $p$ values) is a merger, which would effectively remove one star from the distribution function.
    \item Ejections: a strong (large-angle) scattering between the test star and another stellar-mass object can accelerate it above the local escape speed, $v_{\rm esc}$.  When this happens, the star is removed from the system \citep{LinTremaine80} and can, under certain conditions,  become a hypervelocity star \citep{YuTremaine03, OLearyLoeb08}.  
\end{enumerate}
We will now calculate the different rates of the processes that can remove stars from the distribution. We will begin with an order of magnitude estimate in §\ref{sec:approx} and then present our analytical derivations for strong scattering in §\ref{sec:strong}, star-star interactions in §\ref{sec:star} and star-stellar mass black hole interactions in §\ref{sec:bh} .

\subsection{Approximate Description}
\label{sec:approx}

The rates of processes (i-iv) are governed by basically the same underlying physics, and can be calculated in the usual ``$n-\Sigma-v$'' way as
\begin{equation}
\dot{N} = n \pi p_{\rm c}^2 \left(1+ \frac{2G(M_\star+ \rev{M_2})}{p_{\rm c}v^2} \right) v_\infty,
\end{equation}
where $\dot{N}$ is the per-target rate of close encounters, $p_{\rm c}$ is the critical pericenter for the close encounter (e.g. direct collision, $\mu$-TDE, or tidal capture) to occur, $v_\infty$ is the relative velocity at infinity, $M_\star$ is the mass of the target star, \rev{ $M_{2}$} is the mass of the secondary (another star for star-star encounters or a sBH for TCs/$\mu$-TDEs), and $n$ is the number density of secondaries.

Far from the central MBH, the rates of these close encounters will be set by gravitational focusing, but close to the MBH, encounter rates will be determined by straight-line flyby trajectories.  We will now approximate the rates of close encounters in these two regimes by making the simplified assumption that all stars have the same mass $M_\star$ and are distributed in space with a power-law $n_\star(r) \propto r^{-\gamma_\star}$, and likewise that all sBHs have a common mass $m_{\rm bh}$ and a power-law density profile $n_{\rm bh}(r) \propto r^{-\gamma_{\rm bh}}$.  We further approximate the situation by assuming that all relative velocities are equal to the stellar velocity dispersion, 
\begin{equation}
    \sigma_\star = \sqrt{\frac{GM_{\bullet}}{r(1+\gamma_\star)}},
\end{equation}
sparing us the need to integrate over a relative velocity distribution.  Under these simplifying assumptions, the (per-target) rates of star-star collisions, star-sBH $\mu$-TDEs, and star-sBH tidal captures scale, respectively, as
\begin{align}
    \dot{N}_{\rm coll} \propto \label{eq:NDotCollTwiddle}
    \begin{cases}
    R_\star^2 r^{-\gamma_\star-1/2}, & r < r_{\rm coll}^{\rm tr}  \\
    R_\star M_\star r^{-\gamma_\star+1/2}, & r > r_{\rm coll}^{\rm tr}
    \end{cases} \\
    \dot{N}_{\rm \mu TDE} \propto  \label{eq:NDotMuTDETwiddle}
    \begin{cases}
    R_\star^2 \alpha^{-2/3} r^{-\gamma_{\rm bh}-1/2}, & r < r_{\rm \mu TDE}^{\rm tr}  \\
    R_\star m_{\rm bh} (1+\alpha)\alpha^{-1/3} r^{-\gamma_{\rm bh}+1/2}, & r > r_{\rm \mu TDE}^{\rm tr}
    \end{cases} \\
    \dot{N}_{\rm TC} \propto  \label{eq:NDotTCTwiddle}
    \begin{cases}
    \lambda^2 R_\star^2 \alpha^{-2/3}  r^{-\gamma_{\rm bh}-1/2}, & r < r_{\rm TC}^{\rm tr}  \\
    \lambda R_\star m_{\rm bh} (1+\alpha) \alpha^{-1/3}  r^{-\gamma_{\rm bh}+1/2}, & r > r_{\rm TC}^{\rm tr}.
    \end{cases} 
\end{align}
Here the transition radii between the focused and unfocused interaction regimes are
\begin{align}
    r_{\rm coll}^{\rm tr} =& \frac{R_\star}{4(1+\gamma_\star)} \frac{M_{\bullet}}{M_\star} \approx 2\times 10^{-3}~{\rm pc}~M_6 r_\star m_\star^{-1} \label{eq:rtrans1} \\
    r_{\rm \mu TDE}^{\rm tr} =& \frac{R_\star}{2(1+\gamma_{\rm bh})} \frac{\alpha^{2/3}}{1+\alpha} \frac{M_{\bullet}}{M_\star} \approx 8\times 10^{-4}~{\rm pc}~M_6 r_\star m_\star^{-1} \label{eq:rtrans2} \\
    r_{\rm TC}^{\rm tr} =& \frac{\lambda R_\star}{2(1+\gamma_{\rm bh})} \frac{\alpha^{2/3}}{1+\alpha}  \frac{M_{\bullet}}{M_\star} \approx 2\times 10^{-3}~{\rm pc}~M_6 r_\star m_\star^{-1}, \label{eq:rtrans3}
\end{align}
and we have further defined $\alpha = M_\star/m_{\rm bh}$ as the sBH-star mass ratio, and $\lambda \approx 2$ as the rough enhancement to the $\mu$-TDE cross-section for tidal captures\footnote{In reality, $\lambda \sim \mathcal{O}(1)$ depends weakly on the local properties of the background cluster \citep{LeeOstriker86}.}.  In the approximate equalities above, we have taken $\gamma_\star = \gamma_{\rm bh} = 7/4$, $\lambda = 2$, and $\alpha = 0.1$ for concreteness, and used the shorthand variables $M_6 = M_\bullet / (10^6 M_\odot)$, $m_\star = M_\star / M_\odot$, and $r_\star = R_\star / R_\odot$.

It is important to note that tidal capture becomes impossible when the relative kinetic energy of the target and secondary becomes larger than the maximum amount of mechanical energy that can be injected into the target star's normal modes in a single pericenter passage.  Crudely approximating this condition as $M_\star m_{\rm bh} (M_\star + m_{\rm bh})^{-1} \sigma^2/2 > u_\star GM_\star^2 / R_\star$, where $u_\star <1$ is a fudge factor representing the fraction of the star's binding energy that can be absorbed by mechanical oscillations, we find that tidal capture deactivates at radii below
\begin{equation}
    r_{\rm TC} = \frac{R_\star}{2u_\star (1+\gamma_\star)} \frac{1}{1+\alpha}  \frac{M}{M_\star}.
\end{equation}
Because $r_{\rm TC} \sim r_{\rm TC}^{\rm tr}\alpha^{-2/3}u_\star^{-1}$, and most likely $u_\star \sim \mathcal{O}(0.1)$, $r_{\rm TC}$ is generally greater than $r_{\rm TC}^{\rm tr}$ and tidal captures are always gravitationally focused.

We also want to understand the basic scaling laws describing rates of stellar ejection through strong scatterings.  We approximate the specific impulse delivered to the target star in a strong scattering as $\Delta v \sim (G m_2 / p^2)(p / v_p)$, with $v_p$ the velocity at pericenter.  Angular momentum conservation lets us rewrite this in terms of the velocity ($v_\infty$) and impact parameter ($b$) at infinity: $\Delta v \sim G m_2 / (b v_\infty)$.  Thus the critical impact parameter that allows strong scatterings to eject stars will be $b_{\rm ej} = Gm_2 / (v \Delta v_{\rm ej}$), where $\Delta v_{\rm ej}$ is the critical specific impulse to eject a star.  While it is tempting to equate $\Delta v_{\rm ej} = v_{\rm esc}(r) = \sqrt{2GM/r}$, we must remember that for stars on loss cone orbits, $1-e \ll 1$ and most of the orbit is spent traveling at speeds $v(r)\approx v_{\rm esc}(r)$.  Therefore we approximate $\delta v_{\rm ej} \approx v_{\rm esc}(r) - v(r)$, or
\begin{equation}
    \Delta v_{\rm ej} \approx \frac{1}{2\sqrt{2}} \sqrt{\frac{GMr}{a^2}} 
\end{equation}
for a star on a highly eccentric orbit with $r \ll a$.  We now see that the critical impact parameter for ejection is $b_{\rm ej} \sim 2\sqrt{2} a m_2 / M$.  The local ejection rate from scatterers with number density $n$ will thus be $\dot{N}_{\rm ej} = n \pi b_{\rm ej}^2 v_\infty$, so that
\begin{equation}
    \dot{N}_{\rm ej} \propto a^2 m_2^2 r^{-\gamma - 1/2}. \label{eq:NDotEjTwiddle}
\end{equation}
Here we have left $m_2$ and $\gamma$ general, as they may represent either star-star or star-sBH strong scatterings.  In contrast to Eqs. \ref{eq:NDotCollTwiddle}-\ref{eq:NDotTCTwiddle}, there is no leading-order distinction between gravitationally focused or unfocused regimes for strong scatterings.  Furthermore, there is no leading-order dependence on the mass $m_\star$ of the test star experiencing the scattering (although note that this equation implicitly assumes $m_\star \lesssim m_2$).

While Eqs. \ref{eq:NDotCollTwiddle}-\ref{eq:NDotTCTwiddle} and \ref{eq:NDotEjTwiddle} give local encounter rates $\dot{N}(r)$, we also want to compute their orbit-averaged versions, $\langle \dot{N} \rangle$.  Because of our ultimate interest in the loss cone, we consider highly eccentric orbits with pericenter $p \ll a$, the stellar semimajor axis.  At the order of magnitude level, we can approximate the orbit-averaged rate of some process at a given radius as $\langle \dot{N}(r) \rangle \sim \dot{N}(r) \times (r/a)^{3/2}$, and determine whether the rates are pericenter- or apocenter-dominated to get $\langle \dot{N} \rangle$.

For ejection in strong scatterings, 
\begin{equation}
    \langle \dot{N}_{\rm ej} \rangle \propto a^{1/2} p^{-\gamma + 1} m_2^2, \label{eq:NDotEjTwiddle2}
\end{equation}
so ejection rates will be pericenter-dominated as long as $\gamma > 1$, which we will assume for the remainder of this paper, \rev{as the shallowest profiles are $\gamma \sim$ 1.3 - 1.5 (e.g., \citealt{BW77}).} For other types of close encounters, encounter rates will likewise be pericenter-dominated at small radii $r< r^{\rm tr}$ so long as $\gamma >1$, but the situation at large radii $r > r^{\rm tr}$ is more complicated.  From Eqs. \ref{eq:NDotCollTwiddle}-\ref{eq:NDotTCTwiddle}, we see that close encounter rates will always be pericenter-dominated when $\gamma > 2$, but when $1< \gamma < 2$, the situation depends on whether $p > p^{\rm tr}$, a transitional pericenter\footnote{If $\gamma=2$, then rates are pericenter-dominated for $p<r^{\rm tr}$, but each decade in radius contributes equivalently when $p>r^{\rm tr}$.}.  More specifically, we find that when $1< \gamma < 2$, close encounter rates scale as
\begin{align}
    \langle \dot{N}_{\rm ce}\rangle \propto \label{eq:NDotCETwiddleA}
    \begin{cases}
    p^{-\gamma+1}a^{-3/2}R_{\rm ce}^2, & p < p^{\rm tr}~{\rm or}~a<r^{\rm tr}  \\
     a^{-\gamma+1/2}R_{\rm ce}, & p > p^{\rm tr}~{\rm or}~p > r^{\rm tr}.
    \end{cases}
\end{align}
Here we have approximated the transitional pericenter as 
\begin{equation}
    p^{\rm tr} \sim a \left( \frac{R_{\rm ce}M}{a(M_\star + m_2)} \right)^{1/(\gamma-1)} .
\end{equation}
Note that it \rev{is possible that} $p^{\rm tr}>a$; in this situation, rates are always pericenter-dominated, so even though $p^{\rm tr}$ no longer has a physical meaning, the inequality conditions in Eq. \ref{eq:NDotCETwiddleA} are still accurate.  

Alternatively, when $\gamma >2$,
\begin{align}
    \langle \dot{N}_{\rm ce}\rangle \propto \label{eq:NDotCETwiddleB}
    \begin{cases}
    p^{-\gamma+1}a^{-3/2}R_{\rm ce}^2, & p <r^{\rm tr}  \\
     p^{-\gamma+2}a^{-3/2}R_{\rm ce}, & a > r^{\rm tr}.
    \end{cases}
\end{align}
Because of the many permutations of possibilities, we have not disaggregated collision, $\mu$-TDE, and tidal capture rates in the above formulae, instead referring to a ``close encounter radius'' $R_{\rm ce}$ which will be equal to $R_\star$, $r_{\rm t}$, or $r_{\rm TC}$ for these three processes, respectively.  Likewise, we have denoted a general transition radius $r^{\rm tr}$ differentiating between focused and unfocused local encounter rates; the appropriate transitional radius should be chosen from among Eqs. \ref{eq:rtrans1}-\ref{eq:rtrans3}.

With these scaling relations in hand to provide physical intuition, we now proceed to more rigorously derive ejection and close encounter rates.  

\subsection{Strong Scatterings}
\label{sec:strong}
In this section, we present our analytical derivations of orbit-averaged rates for strong scattering. We employ a stellar distribution function calculated under the assumptions that (i) the stellar density profile $\rho(r) = \rho_{\rm infl} (r/r_{\rm infl})^{-\gamma_\star}$, and (ii) the potential $\psi = GM_{\bullet} /r$ is everywhere Keplerian\footnote{Throughout the remainder of this paper, we adopt the usual stellar dynamics convention of positive-definite potentials and positive-definite energies for bound orbits.}.  Under the assumption of isotropy, we apply an Eddington integral to $\rho$ to obtain
\begin{equation}
    f_\star(\epsilon) = 8^{-1/2}\pi^{-3/2} \frac{\Gamma(\gamma_\star+1)}{\Gamma(\gamma_\star-1/2)} \frac{\rho_{\rm infl}}{\langle m_\star \rangle} \left( \frac{GM_{\bullet}}{r_{\rm infl}}  \right)^{-\gamma_\star} \epsilon^{\gamma_\star-3/2} \label{eq:DF}
\end{equation}
where the (positive-definite) specific orbital energy is, for a given star at radius $r$ and velocity $v$, $\epsilon = \psi(r) - v^2/2$; $\langle m_\star \rangle$ is the average mass in the stellar population; and $r_{\rm infl}$ is the radius that encloses a mass in stars equal to the MBH mass.
\subsubsection{Equal mass scatterer}
We first consider strong scattering from an equal mass scatterer, e.g., from other stars. We will assume that all stars have the same mass $m_\star$ and that the distribution of velocities is isotropic at every point. Let us consider a test star whose velocity is $\bold{V}$. The probability for such a star to have an encounter during a time ${\rm d}t$ that will increase its velocity to  $\bold{V +  \delta v}$ is \citep{Henona}: 

\begin{equation}
P = \frac{8 \pi G^2 m_\star^2 {\rm d}t}{ \delta  v^5} {\rm d} \delta v_x {\rm d} \delta v_y {\rm d} \delta v_z \int_{v_0}^{\infty}f_\star (v ) v  \,{\rm d}v\
\end{equation}
where $f_\star (v)$ is the distribution function of stars and $v_0$ is given by: 
 \begin{equation}
v_0=  \frac{1}{ \delta v} \lvert \bold{V \cdot \delta v} +  \delta v^2 \rvert .
\end{equation}
Let $\theta $  be the angle between $\bold{V}$ and $\bold{\delta v}$, with $v_{\rm esc}$ the escape velocity at this point.  The star will be ejected if : 
\begin{equation}
\label{eq:conditionV1}
V^2 +  \delta v^2 + 2 V \delta v \cos \theta \geq v_{\rm esc}^2.
\end{equation}
Replacing cartesian coordinates by spherical coordinates, the local ejection rate, i.e. the probability for one star to be ejected during a time ${\rm d}t$, is  given by \citep{Henon}:
\begin{equation}
\dot{N}_{\rm ej} = \frac{32 \pi^2 G^2 m_\star^2 }{3 V ( v_{\rm esc}^2-V^2)^2} \int_{ \sqrt{v_{\rm esc}^2-V^2} }^{v_{\rm esc}} (v^2+ V^2-v_{\rm esc}^2)^{3/2} f_\star (v) v  \,{\rm d}v .
\end{equation}
Assuming Keplerian motion and an escape velocity $v_{\rm esc}=\sqrt{2GM_\bullet/r}$, the local ejection rate becomes: 
\begin{equation}
\label{eq:ejectionrate}
 \dot{N}_{\rm ej}= \frac{ 2^{2 - \gamma_\star}  \pi \rho_{\rm infl} a^2 m_\star   V^{1 + 2\gamma_\star}  }{( 1+\gamma_\star) M^2} \left(\frac{GM_\bullet }{r_{\rm infl}} \right)^{ - \gamma_\star}
\end{equation}
with $V$ the local Keplerian velocity of the test star and $a$ its semimajor axis. 
As the relaxation time is much longer than an orbital period, the local ejection rate  per star can then be orbit-averaged \rev{following Eq.\ref{eq:orbitAvg}} as: 

\begin{align}
\label{eq:Orbitaverage}
 \langle \dot{N}_{\rm ej} \rangle= &\frac{2^{2-\gamma_\star} \rho_{\rm infl} m_\star G^{1/2} a ^{3/2- \gamma_\star} r_{\rm infl}^{\gamma_\star}  }{(\gamma_\star +1) M_\bullet ^{3/2}  (1-e^2)^{\gamma_\star-1}}\\
 &\times \int_{0}^{\pi} \frac{(1+e^2+ 2 e \cos \nu )^{1/2+\gamma_\star} }{ (1+ e \cos \nu )^2 }  \,{\rm d}\nu  \notag 
\end{align}

with $\nu$ the true anomaly, and $e$ the eccentricity of the star. 

The orbit-averaged ejection rate does not have a general closed form.  
However, for relevant values of $\gamma_\star$ which are integers or half-integers, an analytical solution exists.  Analytic solutions for some physically motivated values of $\gamma_\star$ are presented in Appendix \ref{app:analytics}.

\subsubsection{Unequal mass scatterer}
Let us now consider strong scattering from an unequal mass scatterer, i.e. stellar mass black holes with $ \alpha =m_\star/m_{\rm bh}$. The probability for a star with velocity  $\bold{V}$ to have an encounter during a time ${\rm d}t$ that will increase its velocity to $\bold{V + \delta v}$ becomes: 
\begin{equation}
P = \frac{8 \pi G^2 m_{\rm bh}^2 dt}{ \delta  v^5} {\rm d} \delta v_x {\rm d} \delta v_y {\rm d} \delta v_z \int_{v_0}^{\infty}f_{\rm bh} (v ) v  {\rm d}v,
\end{equation}
where $f_{\rm bh} (v)$ is the distribution function of stellar mass black holes (also assumed to reflect a power-law density profile, with $\rho_{\rm bh} \propto r^{-\gamma_{\rm bh}}$), and $v_0$ is given by: 
 \begin{equation}
v_0=  \frac{1}{ \delta v} \lvert \bold{V \cdot \delta v} + \frac{m_\star+ m_{\rm bh}}{2 m_{\rm bh}} \delta v^2 \rvert
\end{equation}
The condition for the star to escape remains unchanged. Therefore, after some algebraic manipulations, the local ejection rate becomes : 
\begin{equation}
\begin{split}
\dot{N}_{\rm ej,u} = 16 \pi^2 G^2 m_{\rm bh}^2 \left( \int_{v_1}^{v_2} I_A f_{\rm bh}(v) v {\rm d}v  + \int_{v_2}^{v_3} I_B  f_{\rm bh}(v) v {\rm d}v +\right. \\ \left. \int_{v_3}^{v_{\rm esc}}  I_C f_{\rm bh}(v) v {\rm d}v \right) \label{eq:unequalEjectionsLocal}
\end{split}
\end{equation}
where 
\begin{equation}
\begin{split}
I_{A} &=\frac{2\left[v^{ 2}-\alpha\left(v_{\rm esc}^{2}-V^{2}\right)\right]^{3 / 2}}{3 V\left(v_{\rm esc}^{2}-V^{2}\right)^{2}} \\
I_{B} &=\frac{2\left[v^{ 2}-\alpha\left(v_{\rm esc}^{2}-V^{2}\right)\right]^{3 / 2}+v \left[2 v^{ 2}-3 \alpha\left(v_{\rm esc}^{2}-V^{2}\right)\right]}{6 V\left(v_{\rm esc}^{2}-V^{2}\right)^{2}}\\ 
&-\frac{\left(2 v_{\rm esc}+V\right)}  {6 V\left(v_{\rm esc}+V\right)^{2}}+ \frac{(1+\alpha)^{2}}{8 V\left(V+v\right) } \\
I_{C} &=\frac{3 v_{\rm esc}^{2}-V^{2}}{3\left(v_{\rm esc}^{2}-V^{2}\right)^{2}}+\frac{(1+\alpha)^{2}}{4\left(V^{2}-v^{2 }\right)}
\end{split}
\end{equation}
and the \rev{integration limits} are given by:
\begin{equation}
\begin{split}
v_1 &= \sqrt{ \alpha (v_{\rm esc}^2 - V^2)} \\
v_2 &= \frac{1}{2} [ (1+ \alpha) v_{\rm esc} - (1- \alpha) V ]\\
v_3 &= \frac{1}{2} [ (1+ \alpha) v_{\rm esc} + (1 + \alpha) V ]
\end{split}
\end{equation}
The local ejection rate per star can then be orbit averaged as in Eq. \ref{eq:orbitAvg}.  This orbit-averaged ejection rate does not have a general closed form, although the local ejection rate (Eq. \ref{eq:unequalEjectionsLocal}) can be written in closed form for physically motivated values of $\gamma_{\rm bh}$.

\subsection{Star-star encounters: physical 
\label{sec:star}
collisions and tidal captures}
Tidal captures and physical collisions between stars, although rare in most astrophysical environments, are non-negligible in the extreme densities of galactic nuclei. Assuming that all stars have the same mass $m_\star$, and that the test star has a velocity $\bold{V}$, the local collision rate is given by
\begin{equation}
 \dot{N}_{\rm coll} = \int {\rm d}^3 \bold{v}  f(\bold{v}) \pi b^2 | \bold{V}-  \bold{v} |
\end{equation}
with the gravitationally focused impact parameter $b^2 = r_{\rm coll}^2 + 4 G m r_{\rm coll}/V_0^2$ and $V_0=  | \bold{V}-  \bold{v} |$.  To take into account both collisions and tidal captures from stars we can rewrite the formula as : 
 \begin{equation}
\dot{N}_{\star, \star} = 2\pi^2\int \displaylimits_0^{v_{\rm esc}} \int \displaylimits_0^{\sqrt{v_{\rm esc}^2-v_{\parallel}^2}}
\sqrt{(V-v_{\parallel})^2 + v_{\bot}^2 }\; b_{\rm max}^2 \; f_{\rm bh}(v) v_{\bot} {\rm d}v_{\bot} {\rm d}v_{\parallel}   
\end{equation}
with $b_{\rm max}^2 = r_{\rm max}^2 + 4 G m r_{\rm max}/V_0^2$, where $r_{\rm max} = $ max$[r_{\rm coll}, r_{\rm capt} (V_0, m_{\star})]$ and $r_{\rm capt}=2R_{\rm t,\star}\approx 2R_\star$ is the approximate tidal capture radius we employ for a high-$\sigma$ environment \citep{Stone+17}. 
This formula is only an approximation to a more rigorous calculation of the tidal capture radius, (see e.g., \citealt{LeeOstriker86}), but it will generally be correct to within $\sim 10 \%$  in a high-$\sigma$ galactic nucleus \citep{Generozov+18}.  

Note that tidal capture cannot occur if the relative velocity $V_0 \gtrsim \eta v_\star = \eta \sqrt{Gm_\star / r_\star}$, where $\eta < 1$ is a dimensionless number of uncertain magnitude.  When $V_0 \ll v_\star$, tidal capture can proceed cleanly through the linear excitation of normal modes in the star.  However, when $V_0 \sim v_\star$, the oscillation modes reach nonlinear amplitudes and may begin to dissipate their energy rapidly through nonlinear mode-mode couplings or by steepening into shocks.  Rapid dissipation of the mode mechanical energy can inflate the star and result in its destruction via a sequence of partial $\mu$-TDEs \citep{Kremer+22}. The fudge factor $\eta$ encodes the boundary between successful and unsuccessful captures; for simplicity, in the remainder of this paper we will take $\eta = 0.1$.  Regardless of the true value of $\eta$, it is clear that at small galactocentric radii, stars are tidally disrupted rather than captured. 

The orbit-averaged star-star collision/tidal capture rate $\langle \dot{N}_{\star, \star} \rangle$ are computed following standard orbit-averaging procedures (Eq. \ref{eq:orbitAvg}).


\subsection{Star-sBH encounters: $\mu$-TDEs and tidal captures }
\label{sec:bh}
A star wandering too close to a stellar mass black hole can be tidally captured or tidally disrupted. The characteristic tidal radius is  $  r_{\rm t} = r_\star (m_{\rm bh} /m_\star )^{1/3} $, while the maximum initial pericentre  distance resulting in tidal capture, $r_{capt }$, is approximated as before by $r_{\rm capt} = 2 r_{\rm t}$, with the same caveat that large $V_0$ are treated as $\mu$-TDEs rather than tidal captures.
\\
Considering a test star with velocity $\bold{V}$, its probability to be tidally captured or disrupted by one of the stellar mass black holes is given by: 
\begin{equation}
\dot{N}_{\rm TC} = 2\pi^2\int \displaylimits_0^{v_{\rm esc}} \int \displaylimits_0^{\sqrt{v_{\rm esc}^2-v_{\parallel}^2}}
V_0\; r_o^2 \; [1+ 2 \frac{G (m_\star + m_{bh})}{r_o V_0^2 }] f_{\rm bh}(v) v_{\bot} {\rm d}v_{\bot} {\rm d}v_{\parallel} ,   
\end{equation}
where 
\begin {equation}
\begin{split}
V_0 & = \sqrt{(V-v_{\parallel})^2 + v_{\bot}^2 } \\ 
r_0 & = \max[r_{capt} (V_0,m_{\rm bh}/m_\star ) , r_{\rm t}]
\end{split}
\end {equation}

For practical evaluation, the tidal capture rate is orbit averaged following Eq. \ref{eq:orbitAvg}.


\subsection{Comparison between the different close encounters}

\begin{figure}
     \centering
     \begin{subfigure}[b]{0.5\textwidth}
         \centering
         \includegraphics[width=\textwidth]{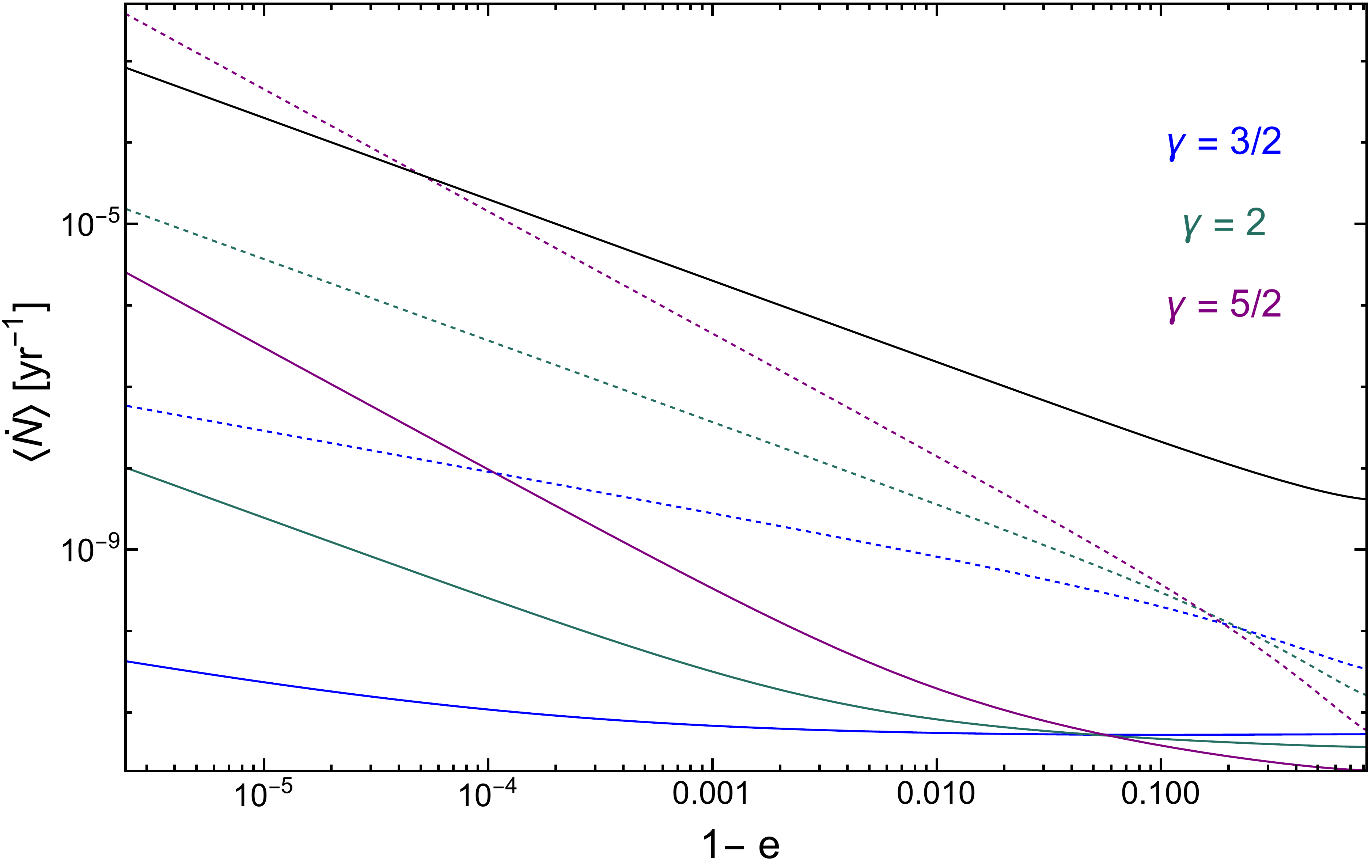}
     \end{subfigure}
     \hfill
     \begin{subfigure}[b]{0.5\textwidth}
         \centering
         \includegraphics[width=\textwidth]{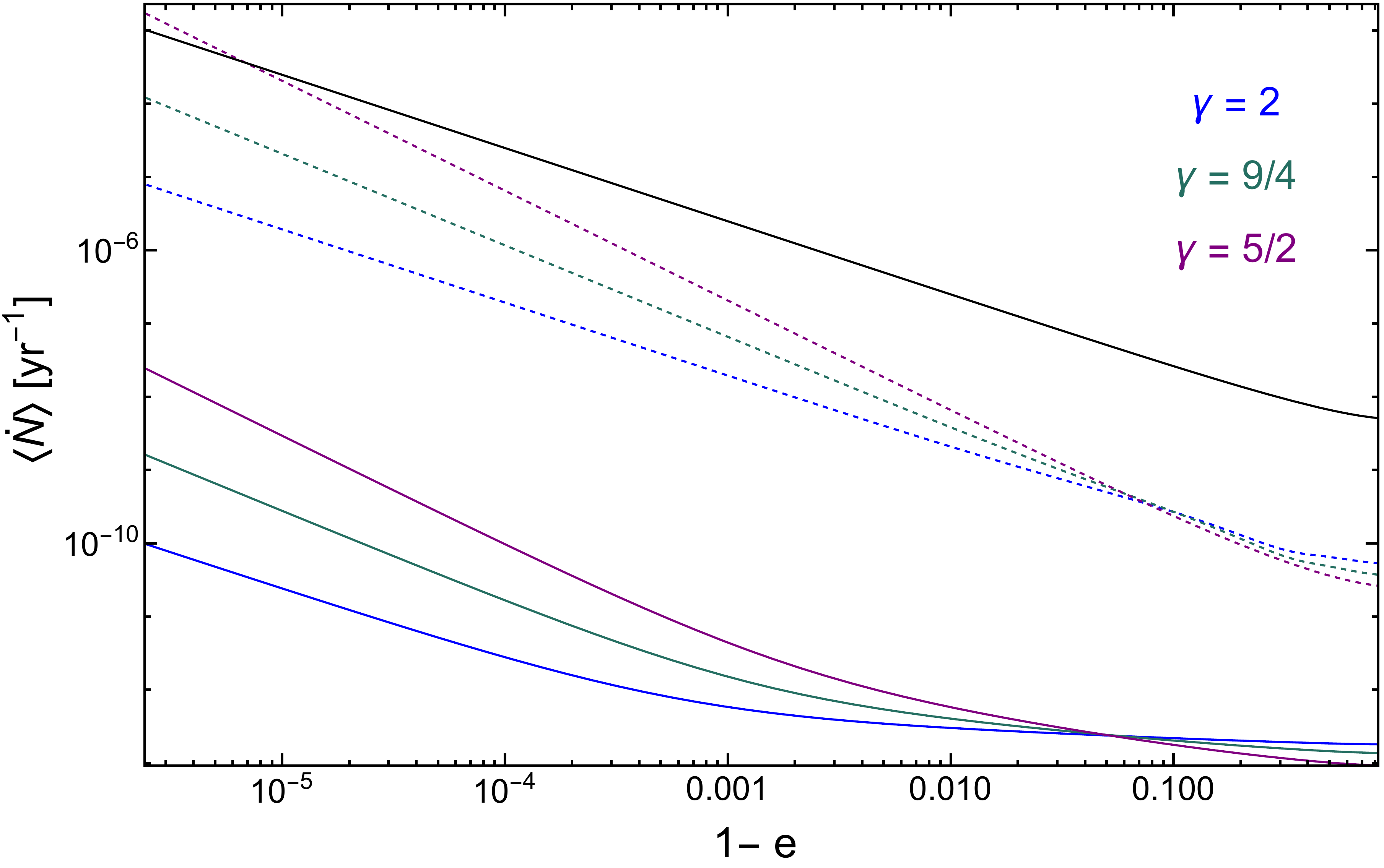}
     \end{subfigure}
        \caption{{\it Top}: Average rates of close encounters between stars, as a function of the dimensionless test star pericenter $1-e$.  The test star has semimajor axis $a=r_{\rm infl}$ and results are shown for different stellar slopes (color-coded).  Dashed lines show strong scattering ejection rates; full lines show tidal captures and stellar collisions. Different colors corresponds to different slopes of the star $\gamma_{\star}$. {\it Bottom}: same as above, but now for unequal-mass encounters between stars and black holes of mass $m_{\rm bh}=30 M_\odot$. The slope of the star is fixed $\gamma_{\star}=3/2$ and the different colors correspond to different slopes of the stellar mass black holes $\gamma_{bh}$. In both panels, the inverse angular momentum relaxation time is plotted in black, $M_{\bullet} = 10^6 M_{\sun}$, $r_{\rm infl}= 0.761$ pc, and $m_{\star} = 1 M_{\sun}$.}
        \label{fig:closeinfluence}
\end{figure}

Given the many close encounters at play in a galactic nucleus, it is useful to compare their relative rates.  We perform this comparison in Fig. \ref{fig:closeinfluence}, which illustrates how strong scattering rates and other close encounter rates vary as one changes stellar eccentricity $e$, and various properties of the central cusp (e.g. $\gamma_\star$, $\gamma_{\rm bh}$).  The rate of angular momentum relaxation, 
\begin{equation}
    \dot{N}_{\rm AM} \sim \mu(\epsilon) (1-e^2) 
\end{equation}
\rev{
is also plotted as a useful reference rate (here $\mu(\epsilon)$ is the diffusion coefficient as computed in Eq. \ref{eq:diffAvg}):}
loss cone shielding is likely to only become important insofar as the rate of a destructive close encounter can become comparable to $\dot{N}_{\rm AM}$. 

We see that in general, $\dot{N}_{\rm AM} \gg \dot{N}_{\rm ej}$ so long as the relevant power-law indices $\gamma \lesssim 2$.  In contrast, $\gamma = 5/2$ leads to ejection rates at large $a$ that can exceed $\dot{N}_{\rm AM}$.  At small $a$, ejection rates decline and can be exceeded by close encounter rates, but regardless of $\gamma$ and $a$, $\dot{N}_{\rm AM} \gg \dot{N}_{\rm ce}$ always.  

To summarize, these simple rate comparisons show that (i) destructive close encounters will never effectively shield the loss cone; (ii) ejections in strong scatterings are likely to shield the loss cone for $a\sim r_{\rm infl}$ and $\gamma > 2$; (iii) at $a \ll r_{\rm infl}$, strong scatterings become subdominant to close encounters as a way to remove stars from the distribution function.  Ultimately, however, these rate comparisons only provide an order of magnitude check of the competition between different physical processes, so we now move on to computing time-dependent solutions to the loss cone problem in the presence of strong scattering and close encounters.

\section{The Shielded Loss Cone}
\label{sec:loss}
\subsection{Loss Cone Theory}
The radial distribution of stars and compact objects evolves over time due to two-body relaxation. In the continuum limit, and assuming spherical symmetry, the stellar population is represented with a distribution function $ f(\epsilon,J)$, where $J$ is the specific angular momentum of a stellar orbit. For the near-radial orbits relevant for TDEs, angular momentum relaxation is much faster than energy relaxation and we will use a two-timescale argument: we separate $f(\epsilon,J) = f_\epsilon(\epsilon) f_j(J)$ and assume a frozen distribution of energy. 
By assumption, stars are fixed in bins of orbital energy, but allowed to diffuse through angular momentum space in a random walk. This process is captured by the orbit-averaged Fokker-Planck equation \citep{MerrittWang05},
\begin{equation}
\frac{\partial f }{\partial \tau} = \frac{1}{4j}\frac{\partial}{\partial j } \left( j\frac{\partial f}{\partial j}\right) \label{eq:FP1}
\end{equation}
where $j \equiv J / J_{c}(\epsilon)=\mathcal{R}^{1 / 2}$ is a dimensionless angular momentum variable (normalized by the angular momentum of a circular orbit, $J_{\rm c}$) and $\tau \equiv \mu(\epsilon) t \approx t/t_{\rm r}$ is a dimensionless version of time $t$, with $\mu(\epsilon)$ the orbit-averaged diffusion coefficient at specific energy $\epsilon$:
\begin{equation}
\mu(\epsilon)=\frac{1}{P(\epsilon)} \oint \frac{d r}{v_{\rm r}} \lim _{\mathcal{R} \rightarrow 0} \frac{\left\langle(\Delta \mathcal{R})^{2}\right\rangle}{2 \mathcal{R}}. \label{eq:diffAvg}
\end{equation}
Here $P(\epsilon)$ is the orbital period of a radial orbit of energy $\epsilon$, $v_{\rm r}$ is the star's radial velocity, and the local diffusion coefficient $\langle (\Delta \mathcal{R} )^2\rangle$ is presented in Appendix \ref{app:diff}. Hence, $\tau \approx t / t_{\rm r}$, with $t_{\rm r}$, the energy relaxation time.

In the presence of strong scatterings and destructive close encounters, stars can be ejected or otherwise removed from the distribution, which we model by adding a sink term to the Fokker-Planck equation: 
\begin{equation}
\frac{\partial f }{\partial \tau} = \frac{1}{4j}\frac{\partial}{\partial j } \left( j\frac{\partial f}{\partial j}\right) -  \frac{\langle \dot{N}_{\rm ej} \rangle}{\mu (\epsilon)}  f. \label{eq:FP2}
\end{equation}
Here $\langle \dot{N}_{\rm ej}\rangle $ is the orbit-averaged rate of close encounters, as derived in section \ref{sec:strong}  for strong scattering and \rev{\ref{sec:star} for star-star encounters and \ref{sec:bh} for star-black hole encounters. }

The intial and inner boundary conditions depend on a dimensionless diffusivity parameter $q(\epsilon)=\mu(\epsilon)P(\epsilon)/j^2_{\rm lc}(\epsilon)$, which is defined in terms of the size of the loss cone in dimensionless angular momentum space ($j_{\rm lc} \approx \sqrt{ 2GM R_{\rm t}}$).  The value of $q$ determines whether the loss cone evolves in the ``empty'' ($q \ll1$; stars immediately destroyed once $j\le j_{\rm lc}$) or ``full'' ($q \gg 1$; stars may move in and out of the loss cone multiple times per orbit) limits. 

When $q(\epsilon) \ll 1$, 
one can assume an absorbing boundary condition at the loss cone, and a zero-flux boundary condition at $j=1$:
\begin{equation}
f\left(j \leq j_{\rm lc}, t\right)=0 ;\left.\quad \frac{\partial f}{\partial j}\right|_{j=1}=0.
\end{equation}
Conversely, when $q(\epsilon) \gg 1$, the distribution function does not go to zero until a much smaller value of dimensionless angular momentum, \rev{$j_0= j_{\rm lc}(\epsilon) \exp(-\alpha/2)$}, where

\begin{equation}
    \rev{\alpha(q) \approx \left(q^2 + q^4 \right)^{1/4}}
\end{equation}
is an approximate flux variable that smoothly bridges the empty and full loss cone limits, \rev{which has been derived by returning to the local (non-orbit-averaged) Fokker–Planck equation and determining how f varies with radial phase assuming $f = 0$ at periapsis \citep{CohnKulsrud78, Merritt13}.} \footnote{\rev{ Here $\alpha$ stems from the equation without a sink term. Hence we checked the impact of a different $j_o$ on the solutions that we derived and found that, as long as $j_o \ll j_{lc}$, the impact was negligible.}} We thus have a more general set of boundary conditions that we use for all numerical solutions:
\begin{equation}
f\left(j \leq j_{\rm 0}, t\right)=0 ;\left.\quad \frac{\partial f}{\partial j}\right|_{j=1}=0.
\end{equation}

In both cases we take as an initial condition an isotropic distribution, $f(j)=1$, for $j_0 \le j \le 1$, and set $f(j)=0$ elsewhere.

The flux of stars that scatter into the loss cone per unit time and energy is given by:
\begin{equation}
\mathcal{F}(t ; \epsilon)= 2 \pi^2 \mu(\epsilon) P(\epsilon) J_{\rm c}^2(\epsilon) f_\epsilon(\epsilon) \left ( j\frac{\partial f_j(j,t) }{\partial j}\right)_{j= j_{\mathrm{lc}}}. \label{eq:flux}
\end{equation}

Then, the TDE rate is obtained by by integrating $\mathcal{F}(\epsilon)$ across many bins of energy $\epsilon$, such as : 
\begin{equation}
    \dot{N}_{\rm TDE} (t)  = \int \mathcal{F}(t; \epsilon) {\rm d}\epsilon. \label{eq:totalRate}
\end{equation}

\subsection{Numerical Results}

\begin{figure}
     \centering
     \begin{subfigure}[b]{0.5\textwidth}
         \centering
         \includegraphics[width=\textwidth]{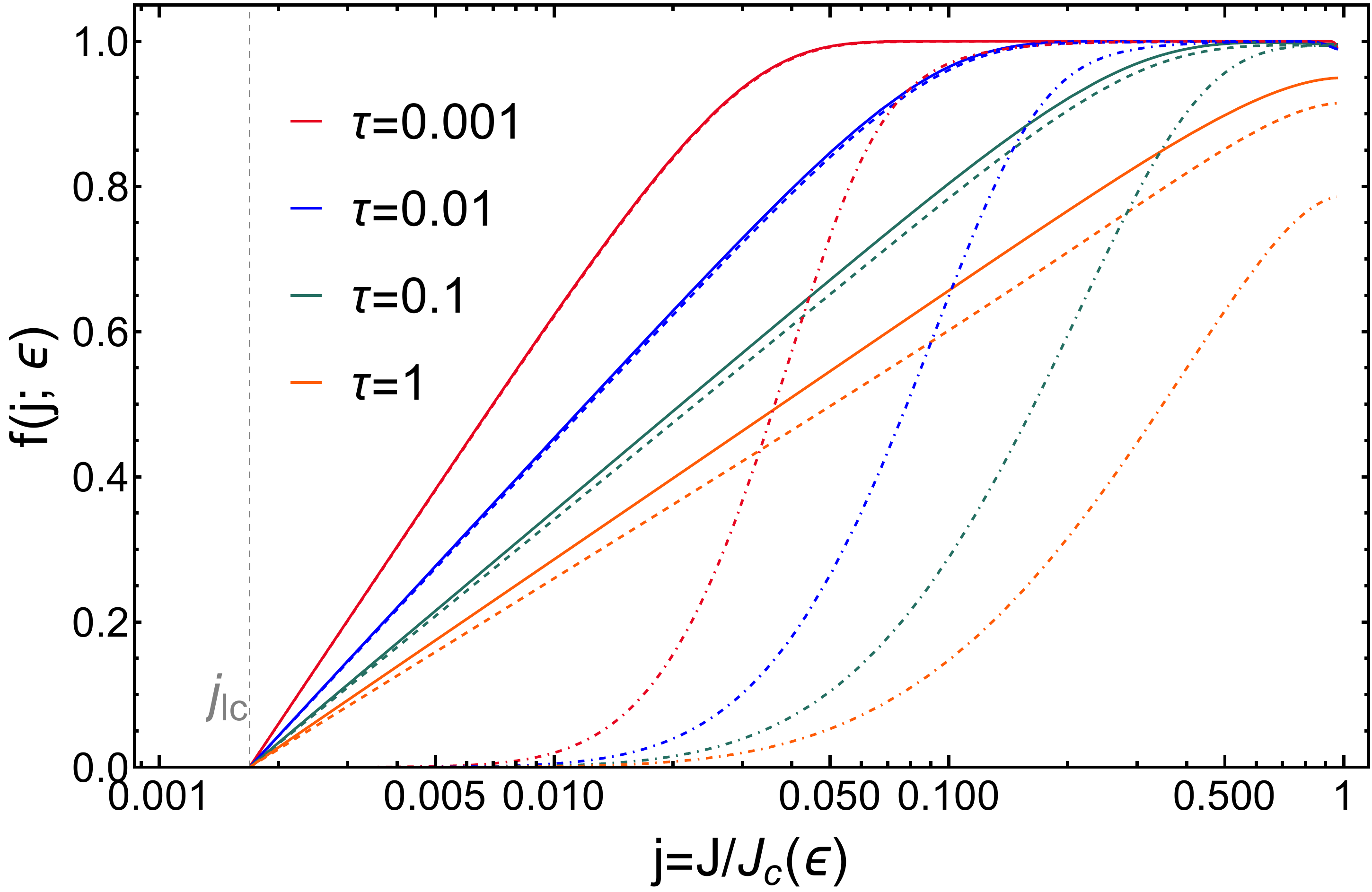}
     \end{subfigure}
     \hfill
     \begin{subfigure}[b]{0.5\textwidth}
         \centering
         \includegraphics[width=\textwidth]{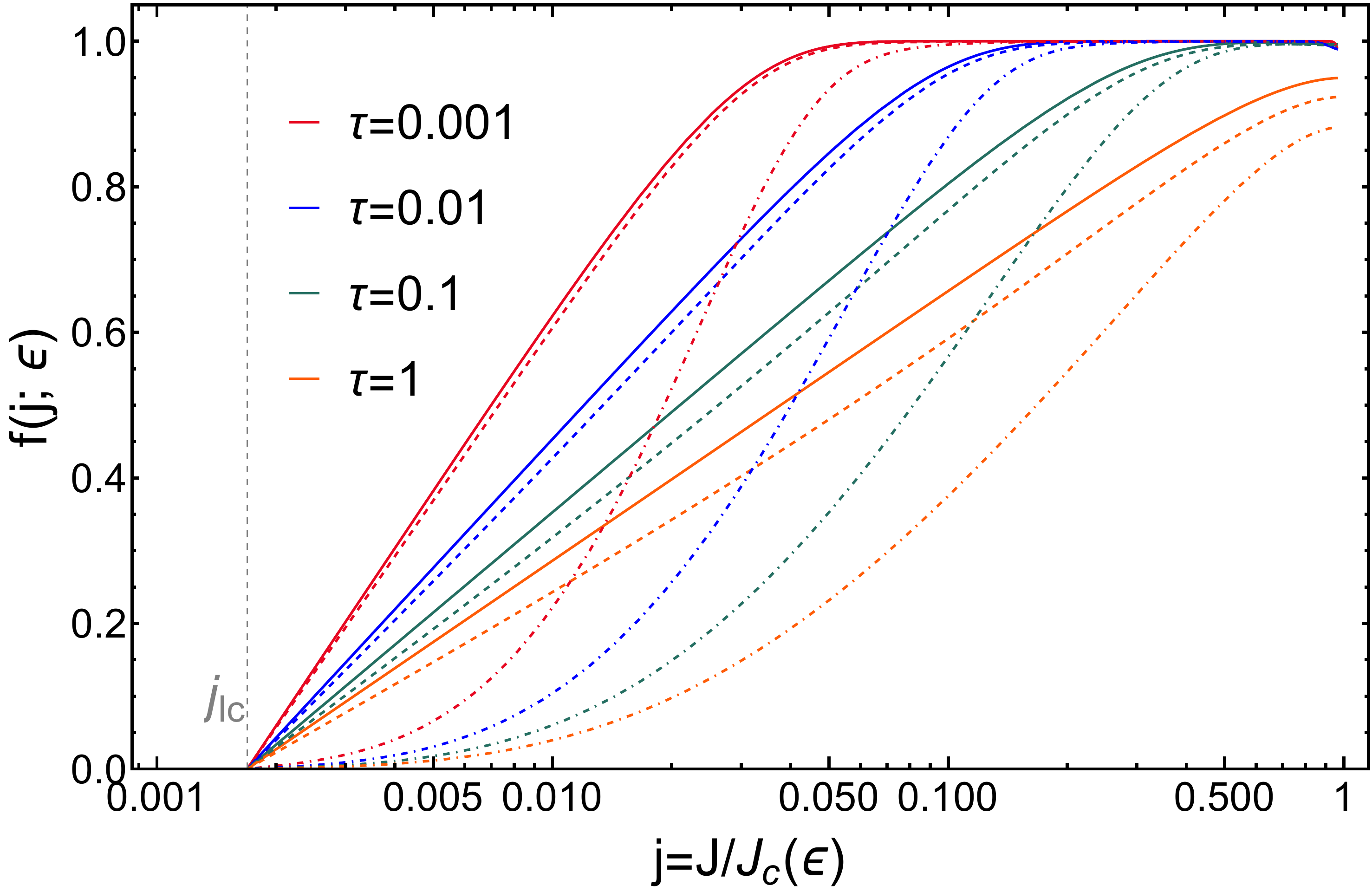}
     \end{subfigure}
        \caption{Numerical evolution of the distribution function $f (j;\epsilon)$ as a function of the dimensionless angular momentum $j$ at fixed energy $\epsilon$ corresponding to $a = r _{inf}$, shown for different snapshots in dimensionless time, $\tau$, for a Milky Way-like galaxy ($M=4\times 10^6 M_\odot$). The solid lines correspond to the evolution without strong scattering or other sink terms (Eq. \ref{eq:FP1}). \textit{Top}: strong scattering (Eq. \ref{eq:FP2}) from stars only; dashed lines show $\gamma_{\star} = 3/2$, dot-dashed lines $\gamma_{\star} = 5/2$. \textit{Bottom}: strong scattering from black holes; dashed lines show $\gamma_{\rm bh} = 2$, dot-dashed lines $\gamma_{\rm bh} = 5/2$. In this panel $\gamma_{\star} =3/2$, $N_{\rm bh}=10^3$, and $m_{\rm bh}= 30 M_{\odot}$. In both panels we see dramatic changes to the distribution $f$ for ultra-steep density cusps, but only minor changes for more shallow profiles.
        }
        \label{fig:FK}
\end{figure}

We begin by solving the modified Fokker-Planck equation (Eq. \ref{eq:FP2}) numerically.  An example of these time-dependent solutions is presented in Fig. \ref{fig:FK}, where we show the evolution of the angular momentum distribution (in a bin of fixed $\epsilon$), both with and without strong scatterings, \rev{in the diffusive regime ($q = 10^{-2}$)}. In the absence of strong scatterings, we see that we quickly achieve the logarithmic Cohn-Kulsrud profile \citep{CohnKulsrud78}.  However, when strong scatterings are turned on, a range of outcomes are possible. For ($\gamma_\star = 3/2$ and $\gamma_{\rm bh}=2$), both stars and sBHs have a very small impact at early times and a moderate one at later times. \rev{The distribution function still has a very similar form to the logarithmic Cohn- Kulsrud solution, albeit with a small depletion almost evenly distributed across angular momentum. For $\gamma_{\rm bh} = 5/2$ (or an  ultra-steep density star profiles $\gamma_\star=5/2$), the distribution function} can be radically modified relative to the standard Cohn-Kulsrud solution, with a major depletion of low-$j$ orbits.  The ``cavity'' carved out of the distribution function at low $j$ in this case reflects the impact of strong scattering, which preferentially ejects stars on the most radial orbits (as it is generally the case that ejection rates for an individual star are pericenter-dominated, see e.g. Eq. \ref{eq:Orbitaverage}).

\begin{figure}
    	\centerline{\includegraphics[width=90mm]{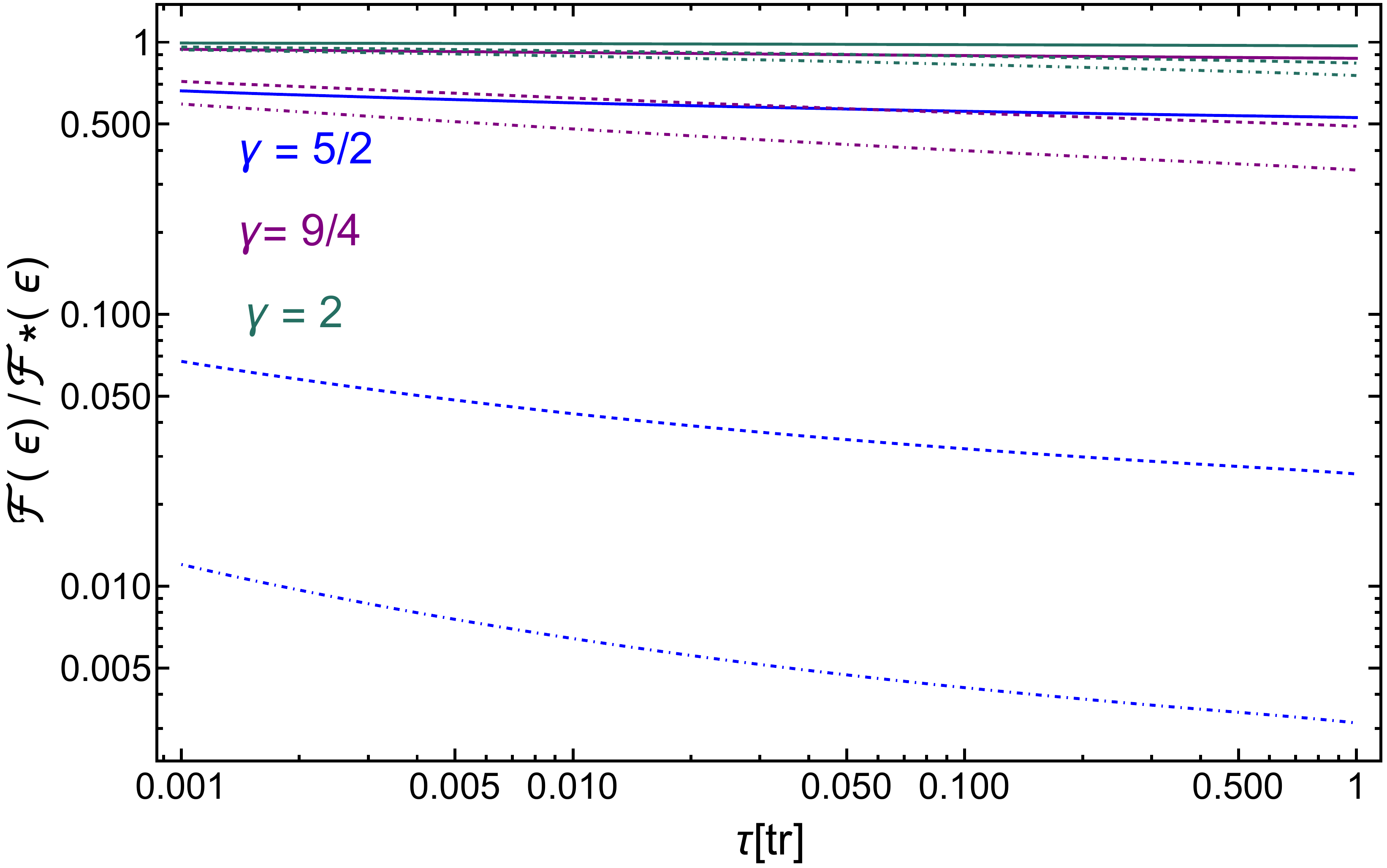}}
        \caption{Evolution of the flux of stars inside the loss cone with strong scattering ($\mathcal{F}$) divided by the flux of stars neglecting strong scattering ($\mathcal{F}_{\rm w}$) as a function of dimensionless time $\tau$ at fixed energy $\epsilon$ corresponding to $a = r _{inf}$, for a Milky Way like galaxy with  $m_{\star} = 1 $, $\gamma_{\star} =3/2$, $N_{\rm bh}=10^3$.  Different colors correspond to different stellar mass black hole density slopes $\gamma_{\rm bh}$ (see legends). Full lines show $m_{\rm bh}= 10 M_{\odot}$, dashed lines show $m_{\rm bh}= 30 M_{\odot}$, and dot-dashed lines $m_{\rm bh}= 50 M_{\odot}$ stars. }
        \label{fig:Flux}
\end{figure}

When strong scatterings are capable of strongly depleting the low-$j$ end of the distribution function, the result is a major reduction in loss cone flux (as $\mathcal{F} \propto (\partial f_j / \partial j)|_{\rm j=j_{\rm lc}}$).  We further explore this effect by computing flux reduction for different black hole slopes and black hole mass\rev{,}  applying Eq. \ref{eq:flux} to our numerical solutions, and plot the results in Fig. \ref{fig:Flux}.  In this figure, we see that in all cases, there is relatively little evolution of the flux suppression factor $\mathcal{F}/\mathcal{F}_{\rm w}$ (here $\mathcal{F}_{\rm w}$ is the flux into the loss cone that would be achieved without the presence of strong scatterings).  Suppression factors $\mathcal{F}/\mathcal{F}_{\rm w}$ are modest $\approx 1$ for $\gamma_{\rm bh}=2$, reach the factor-of-2 level for $\gamma_{\rm bh}=9/4$, and can produce multiple-order-of-magnitude drops in the TDE rate for steeper $\gamma_{\rm bh}$ values.  As shown in  Eq. \ref{eq:unequalEjectionsLocal}, we see stronger levels of flux suppression for larger values of $m_{\rm bh}$.  


\subsection{Analytical solutions of Fokker-Planck with strong scattering}
\label{sec:analytical}
The flux of stars inside the loss cone $\mathcal{F} (\epsilon ) $ is a sharply peaked function, and most of the stars come from energies close to the influence radius. As we have shown, in section \ref{sec:close}, the dominant close-encounter mechanism  at influence radius is strong scattering. Therefore, we searched for an analytical solution to the Fokker-Planck equation with a sink term corresponding to strong scattering (Eq. \ref{eq:FP2}).

\subsubsection{Equal mass scatterer}
\begin{figure}
	\centerline{\includegraphics[width=90mm]{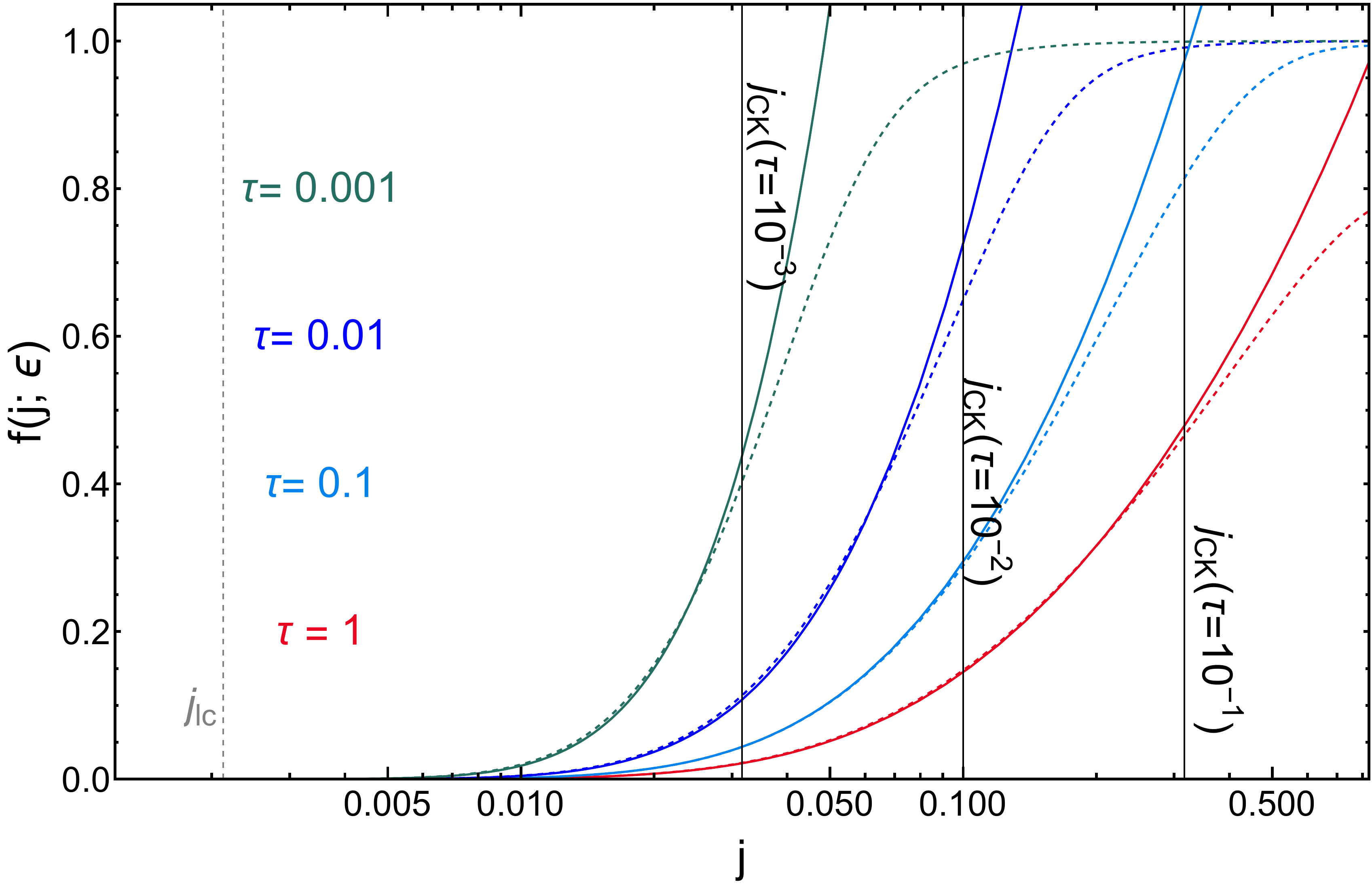}}
    \caption{Evolution of the distribution function $f(j;\epsilon)$ as a function of the dimensionless angular momentum $j$ at fixed energy $\epsilon$ shown for different snapshots in dimensionless time, $\tau$ \rev{in the diffusive regime ($q=10^{-2}$)}.  Here $\gamma_\star = 5/2$ and we consider a single-mass stellar population. The dashed curves show numerical solutions of the 1D Fokker-Planck equation and solid lines show the approximate analytic solution obtained from the method of Frobenius in Eq. \ref{eq:frobenius5.2}.  The agreement is excellent for values of $j \lesssim j_{\rm CK}$, the ``Cohn-Kulsrud'' angular momentum below which the distribution function has had time to relax into a QSS.}
   \label{fig:analytical52equal}
\end{figure}

For an equal mass scatterer, we found an exact analytical form for the orbit-averaged ejection rate for all relevant integer or half-integer values of $\gamma_\star$. The formulas for $\gamma_\star=3/2, 5/2$ can be found in Appendix \ref{app:analytics}.\footnote{Formulas for $\gamma_\star=2$ are in close form with Hypergeometrics series} \rev{We assume a slow variation with respect to $\tau$ such as df/d$\tau \approx$ 0,  and use the method of Frobenius to derive an analytic solution in angular momentum to the modified Fokker-Planck equation for these two power-law distributions.} For the shallower slope, $\gamma_\star= 3/2$, the solution is very close to the Cohn-Kulsrud profile, whereas for the steeper slope of $\gamma_\star= 5/2$, it has an exponential behaviour. 
The solutions for $\gamma_\star= 3/2$ and $\gamma_\star= 5/2$ can be written explicitly as: 
\begin{equation}
\begin{split}
f_{3/2}(j) & = ( 1 + 16 A j) ( a + b \ln{j}) - 32 A b j 
\\
f_{5/2}(j) &=  e^{\frac{- 4 \sqrt{2 A}}{\sqrt{j}}} j^{1/4} [a g_{-}(j) + b e^{\frac{8 \sqrt{2 A}}{\sqrt{j}}} g_{+} (j) ]  
\label{eq:frobenius5.2}
 \end{split} 
\end{equation}
where $A =  \langle \dot{N}_{ej} \rangle/ \mu (\epsilon)$ is a per-star sink term, $\langle \dot{N}_{ej} \rangle$ can be found in Appendix \ref{app:analytics}, \rev{$ g_{\pm} = 1 \pm \frac{\sqrt{j}}{32 \sqrt{ 2 A} } + 0.0022 \frac{j}{ A}$ }and two undetermined constants -- $b$ and $a$ -- need to be found.  These solutions are valid at small $j$, which is the relevant parameter space for stars that could undergo a disruption.  Their explicit time-dependence comes from the undetermined constants $a$ and $b$.  In practice, $b$ is deterministically set as a function of $a$ using the absorbing boundary condition at $j=j_{o}$, so there is only one true time-dependent free parameter.  We find that $a$ depends on the dimensionless time $\tau$ as $a \sim \tau^{-1/2}$. 
Since the approximate time-scale for angular momentum relaxation to occur is $t_j(j)\sim j^2 t_{\rm r} $, our solution is accurate for $ j \lesssim j_{\rm CK} = \sqrt{\tau}$. In Fig. \ref{fig:analytical52equal}, the numerical solutions of the Fokker-Planck equation with strong scatterings (Eq. \ref{eq:FP2}) are compared to our analytical solution ($\gamma_\star=5/2$) at different dimensionless times. The agreement is excellent for values of $j \lesssim j_{\rm CK}$, the ``Cohn-Kulsrud'' angular momentum below which the distribution function has had time to relax into a QSS. 

\subsubsection{Unequal mass scatterer}
For an unequal mass scatterer, we found close form solutions for the local rates and derived approximate solutions for the orbit-averaged rates. We then repeat our application of the Frobenius method to find time-dependent solutions for physically motivated slopes. Specifically, we find solutions for the time evolution of the distribution function of stars scattering off a population of sBHs with power-law slopes of indices $\gamma_{\rm bh}=3/2 ,7/4 , 2, 9/4 $ and $5/2$. The resulting closed form solutions are as follows: 
\begin{equation}
\begin{split}
f_{3/2,u}(j)&= ( 1 + 4 A j) (a + b \ln{j}) - 8 A j b 
\\
 f_{7/4,u}(j)&= (1 + 8 A \sqrt{j})^2 (a - \frac{b}{2}  (2 \ln{16 A} + \ln{j}))  
 \\
- & 2 b (\gamma_{E} + 16 A (-1 + \gamma_{E}) \sqrt{j} +  32 A^2 (-3 + 2 \gamma_{E}) j)  
 \\
f_{2,u}(j) & = j^{-2 \sqrt{A}} (j^{4  \sqrt{A}} a +b)
\\
f_{9/4,u} & = e^{-8 \sqrt{A} j^{-1/4}} j^{1/8} \left( \frac{ \sqrt{\pi b }} {2 A^{1/4}}  (-1 + \frac{ j^{1/4} }{ 64 A^{1/2}} ) \right.
\\
+ &  \left. \frac{a}{ 2 \sqrt{2 \pi}} \cosh{(8  \sqrt{A} j^{-1/4} ) } \right) 
\\
f_{5/2,u}(j)&=  e^{\frac{-4 \sqrt{A}}{\sqrt{j} }} j^{1/4} [a h_{-}(j) + b e^ { \frac{8 \sqrt{A}}{\sqrt{j} }} h_{-} (j)]
\end{split}
\label{eq:frobeniusunequal}
\end{equation}
where $\gamma_E$ is Euler's constant and $h_{\pm} = 1 \pm \frac{\sqrt{j}}{32 \sqrt{A} } + 0.0044 \frac{j}{ A} $.
\\
\begin{figure}
	\centerline{\includegraphics[width=90mm]{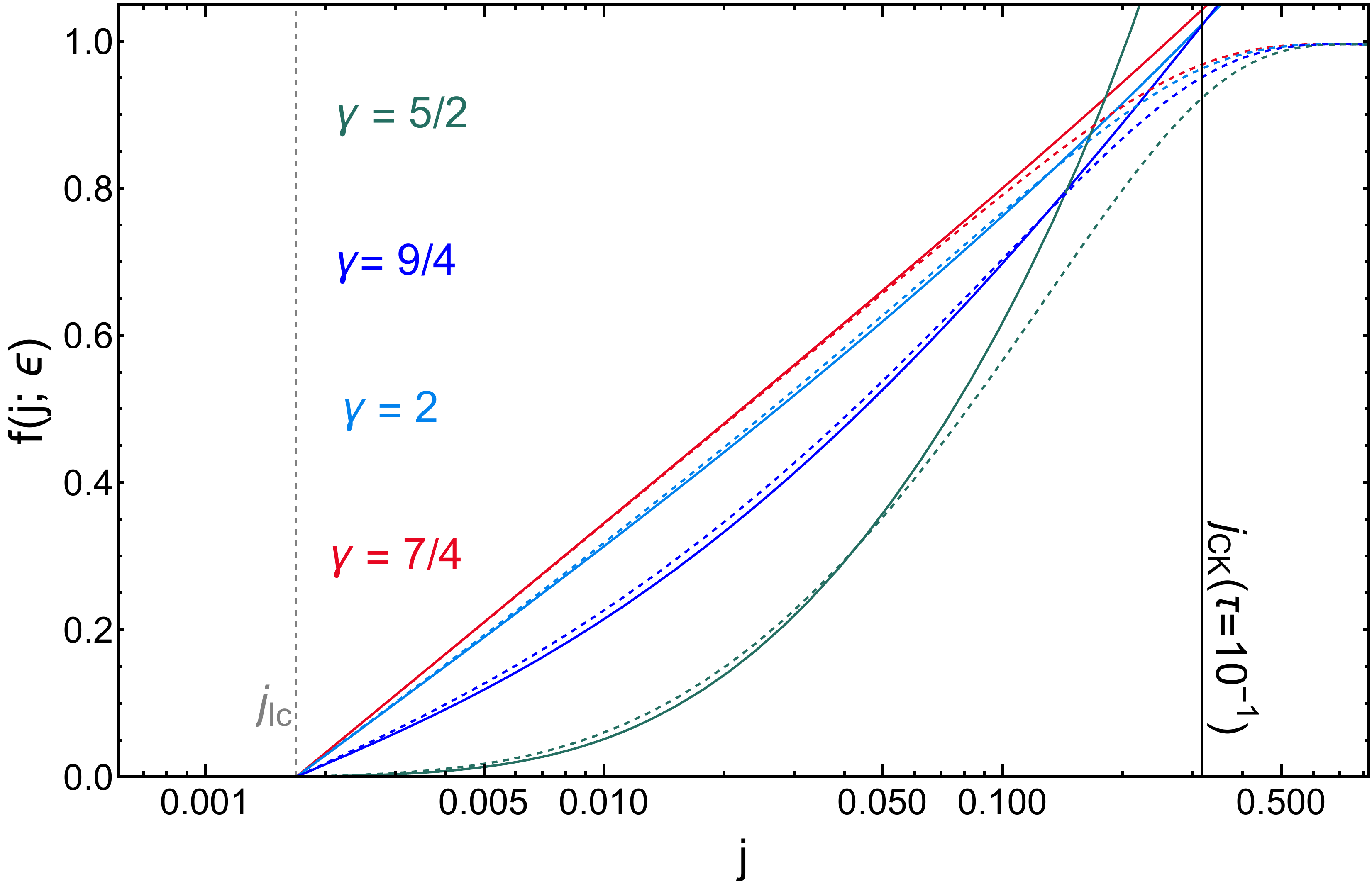}}
    \caption{Evolution of the distribution function $f(j;\epsilon)$ as a function of the dimensionless angular momentum $j$ at fixed energy $\epsilon$, at  dimensionless time, $\tau =0.1$ for different slopes of the black holes \rev{in the diffusive regime ($q=10^{-2}$)}. Here $\gamma_\star = 3/2$, the stellar mass black holes have a mass $m = 30 M_\odot$, and different colors correspond to different sBH slopes $\gamma_{\rm bh}$ (see labels in figure). The dashed curves show numerical solutions of the 1D Fokker-Planck equation and solid lines show the approximate analytic solution obtained from the method of Frobenius in Eq. \ref{eq:frobeniusunequal}.  As in Fig. \ref{fig:analytical52equal}, the agreement is again excellent for values of $j \lesssim j_{\rm CK}$.}
   \label{fig:analyticalunequal}
\end{figure}
As before, explicit time-dependence enters through two undetermined coefficients, $a$ and $b$, although application of the absorbing inner boundary condition allows us to eliminate one of these.  We find that the remaining coefficient can be determined quite accurately with a matching procedure: we set the value of our solution, $f_{\gamma_\star, u}(x j_{\rm cK})$ equal to the value of the Cohn-Kulsrud distribution at $j=x j_{\rm CK}$.  Here $x$ is a dimensionless number of order unity; we find that $x \sim 0.1$ yields a good match.

In Fig. \ref{fig:analyticalunequal}, we compare the analytical solutions we obtained for unequal-mass strong scatterings at various $\gamma_{\rm bh}$ with the numerical solutions to Eq. \ref{eq:FP2}.  In all cases (analytic and numerical) the comparison is done at a fixed dimensionless time $\tau = 0.1 $. As can be seen, the agreement between analytical and numerical solutions is excellent for $ j \lesssim \sqrt{\tau}$.  

\section{Impact on TDE rates}
\label{sec:results}

\subsection{Rates}
As we have seen in Fig. \ref{fig:Flux}, the flux inside the loss cone reaches a quasi-stationary state after a time $t \sim 0.1 \: t_{r}$, with $t_r$, the relaxation time. Therefore, in this section we explore the impact on TDE rates of strong scattering after a time where the quasi-stationary state has been reached. To compute numerical fluxes we compute the diffusion coefficients $\mu (\epsilon)$ for a grid of energies. Equation \ref{eq:FP2} is then integrated forward for the appropriate dimensionless time at each $\epsilon$-value and the flux computed from equation \ref{eq:flux}. Finally, TDE rates are obtained by integrating across the energy bins, equation \ref{eq:totalRate}. To compute analytical fluxes, we use our solutions presented in Sec \ref{sec:analytical} for the appropriate dimensionless time at each $\epsilon$-value, then compute the fluxes with equation \ref{eq:flux} and finally integrate across multiple energy bins to obtain TDE rates, equation  \ref{eq:totalRate}.

\begin{figure}
	\centerline{\includegraphics[width=90mm]{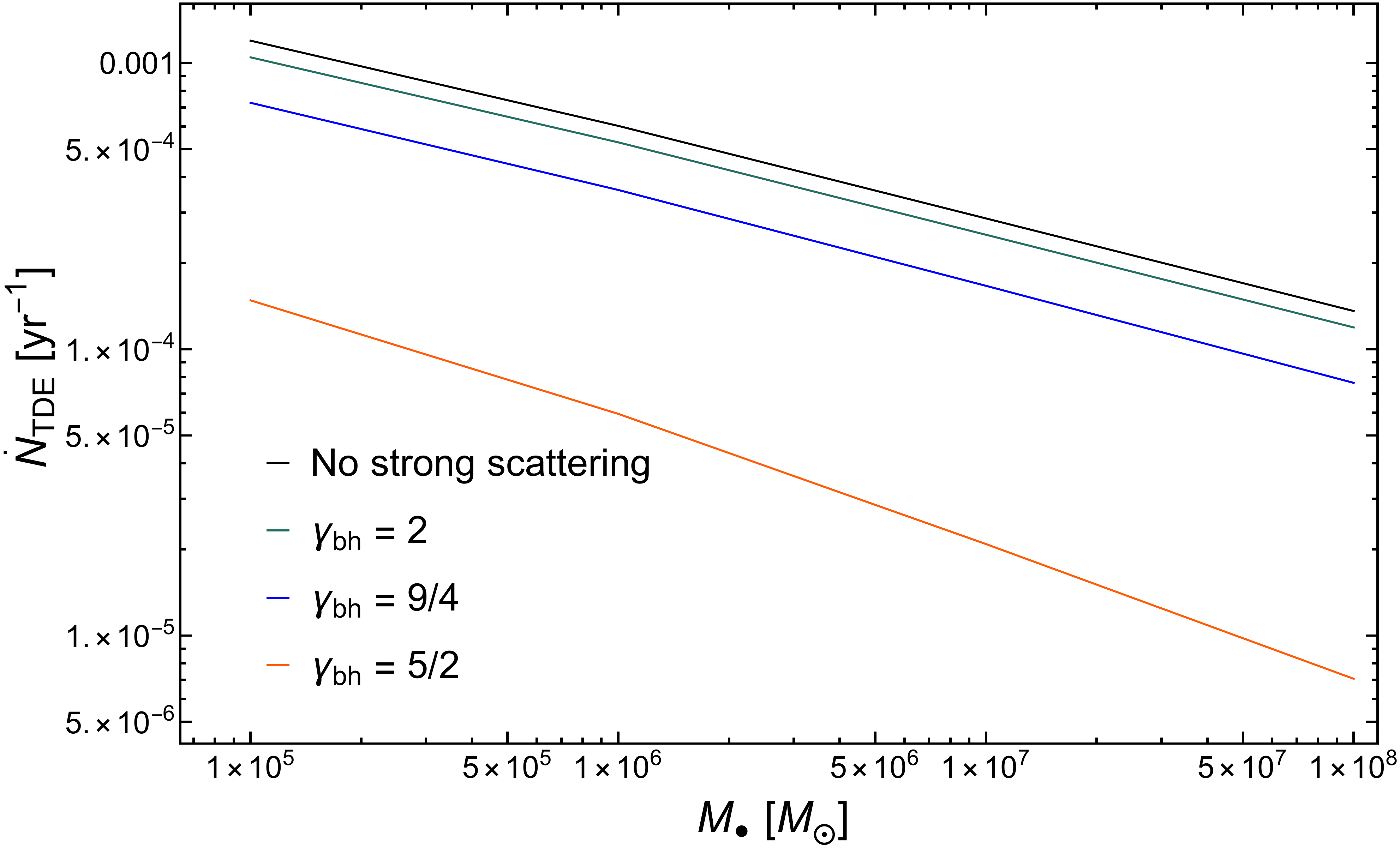}}
    \caption{Tidal disruption rates $\dot{N}_{\rm TDE}$ measured in units of stars per year with (colored curves) and without (black curve) strong scattering. Results are plotted against MBH mass $M_{\bullet}$, with the following parameters:  $m_{\star} = 1 M_{\sun}$, $\gamma_{\star}= 3/2$, $m_{\rm bh} = 30  M_{\sun} $ and different sBH density profile slopes: in particular, we show $\gamma_{\rm bh}=2$, $\gamma_{\rm bh}=9/4$, and $\gamma_{\rm bh}=5/2$ as green, blue, and orange curves, respectively.  Reductions in total TDE rates begin to become significant for $\gamma_{\rm bh} \ge 9/4$.}
   \label{fig:rateSMBH}
\end{figure}

In Fig. \ref{fig:rateSMBH}, we show energy-integrated TDE rates as a function of MBH mass both with and without the effects of strong scattering for cusp galaxies with radius of influence $  r_{\rm infl} = 11~{\rm pc} \left(\frac{M_\bullet}{10^8 M_\odot} \right)^{0.58}$ \citep{StoneMetzger16}.

\begin{figure}
	\centerline{\includegraphics[width=90mm]{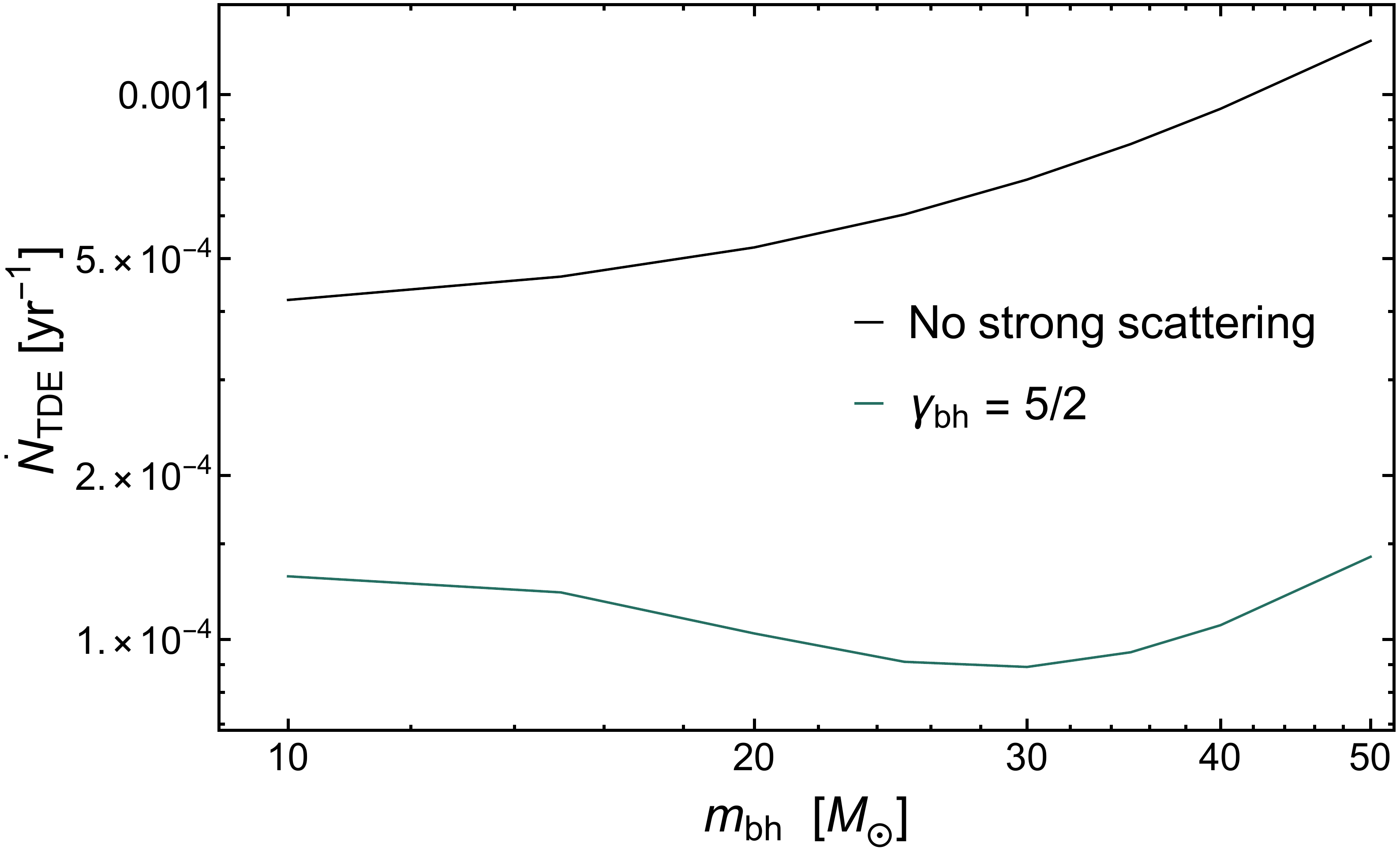}}
    \caption{Tidal disruption rates $\dot{N}_{\rm TDE}$ measured in units of stars per year with (colored curves) and without (black curve) strong scattering as a function of stellar black hole mass $m_{bh}$. Higher stellar mass black holes increase the diffusion coefficients and hence the TDE rates without strong scattering. However, with strong scattering there is a balance between the effect on the diffusion and the higher efficiency at rmoving of stars of heavier sBH. }
   \label{fig:mbhMasses}
\end{figure}

We can note that our TDE rates without strong scattering shown in Fig. \ref{fig:rateSMBH} are a factor of a few higher than in previous loss cone calculations (e.g., \citealt{WangMerritt04, StoneMetzger16}).  This effect is due to the inclusion of stellar mass black holes. Indeed, stellar mass black holes dominate the total relaxation rate and increase diffusion coefficients hence increasing the TDE rates by a factor of a few. This effect is further explored in Fig. \ref{fig:mbhMasses} where we show the influence of different sBH masses on TDE rates both with and without strong scattering.

We see here that the effects of strong scattering are quite modest for $\gamma_{\rm bh}=2$ (reducing TDE rates by $\lesssim 50\%$), rise to a factor of 2 reduction for $\gamma=9/4$, and then become very large (at least an order of magnitude reduction) for $\gamma=5/2$. It is important to note, {\it that the TDE rates we obtain with $\gamma_{bh} =5/2$ are in agreement with the observed TDE rates}. In this example we have taken $m_{\rm bh}=30 M_\odot$. The impact of  different SBH masses is explored in Fig. \ref{fig:mbhMasses}.

 In Fig. \ref{fig:mbhMasses},  we show that in the absence of strong scattering, TDE rates increase by a factor $\approx 3$ as $m_{\rm bh}$ increases from $10M_\odot$ to $50 M_\odot$.  In the presence of strong scattering, however, the TDE rate is substantially lower at all values of sBH mass and is relatively independent of $m_{\rm bh}$.  Although larger $m_{\rm bh}$ values still increase diffusion coefficients, this is offset by the greater rate at which stars are ejected in close encounters (note that both effects scale $\propto m_{\rm bh}^2$).

In these calculations, we considered a broad range of sBH masses, from $m_{\rm bh} = 10 M_\odot - 50 M_\odot$.  Here we are motivated partially by the ``initial mass function'' for sBHs formed via the deaths of massive stars.  Depending on the metallicity of the stellar population, the maximum masses for sBHs formed in this way is likely to be $\approx 50 M_\odot$ \citep{SperaMapelli17}, although even with a favorable metallicity, the large majority of sBHs will have substantially lower masses\footnote{We note that this type of calculation is quite sensitive to uncertain prescriptions for line-driven wind mass loss \citep{Belczynski+10}.}. \rev{However, more recent  simulations have shown that $30 M_{\sun}$ can be formed at stellar metallicity \citep{Bavera}. }  In addition, once a population of sBHs is situated in a galactic center environment, it may be able to gain mass through mergers and accretion.  Most notably, any episode of large-scale MBH accretion will grind down and capture many of the pre-existing sBHs into the MBH accretion disk \citep{SyerRees91}; once embedded in this disk, they can be processed to larger masses either via mergers \citep{Bellovary+16, Tagawa+20} or from direct gas accretion \citep{GilbaumStone22}.  After the AGN episode ends, the population may relax into a more spherical configuration of the type we assume here \citep[although see][]{SzolgyenKocsis18}.  Efficient sBH growth through merger may also occur during periods of time when a dense nuclear star cluster lacks a MBH \citep{AntoniniRasio16}, either early during its evolution or at later points if an MBH is temporarily ejected via gravitational wave recoil \citep{GualandrisMerritt08}.

The rate suppression due to strong scatterings can depend strongly on the semimajor axis, or equivalently the energy $\epsilon$, of the stars in question.  At low values of $\epsilon$, TDE rates can be suppressed by one or more orders of magnitude in the presence of high-mass and steeply cusped sBHs, but the suppression is often milder (especially at high $\gamma_{\rm bh}$) for the subset of stars on tightly bound, high-$\epsilon$ orbits.  We have seen this numerically, and established it analytically  in Eqs. \ref{eq:ejectionrate} and \ref{eq:Orbitaverage}.  A consequence of this is that galactic nuclei can see a relative suppression of high-$\beta$ TDEs, which are overwhelmingly produced in the full loss cone (or ``pinhole'') regime where rates are suppressed most heavily by strong scatterings.  Although the observational implications of high- vs low-$\beta$ TDEs are not yet clear, high-$\beta$ events may exhibit faster circularization \citep{Hayasaki+13, Bonnerot+16}, produce stronger thermal soft X-ray emission \citep{Dai+15}, produce stronger sub-relativistic outflows from the circularization process \citep{MetzgerStone16, LuBonnerot20}. These observational signatures could be less frequently produced in galaxies with large and steeply cusped populations of sBHs.

\begin{figure}
	\centerline{\includegraphics[width=90mm]{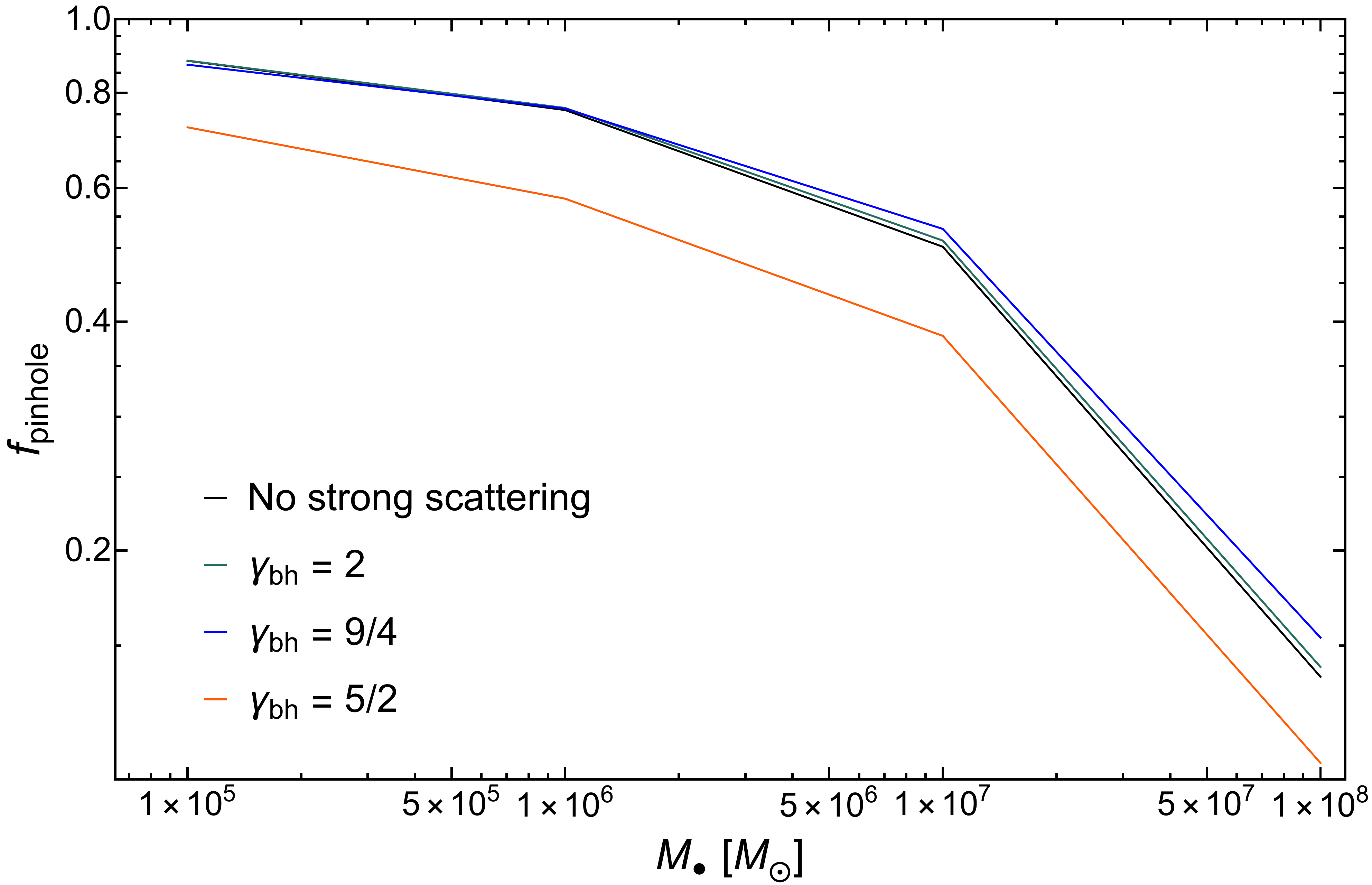}}
    \caption{The fraction, $f_{\rm pinhole}$, of all TDEs in a galaxy which fall into the pinhole regime of tidal disruption, plotted against MBH $M_{\bullet}$.  The black curve shows this dependence without strong scattering, while the colored curves show the pinhole fraction in the presence of strong scattering (see labels in-panel). TDEs in the pinhole (or full loss cone) regime can access any $\beta$, while TDEs in the opposite, diffusive, regime, almost always have $\beta \approx 1$.}
   \label{fig:pinhole}
\end{figure}

We quantify this effect in Fig. \ref{fig:pinhole}, which shows $f_{\rm pinhole}$, the fraction of TDEs occuring in the pinhole regime, as a function of MBH mass $M_\bullet$ (with and without strong scattering). Without strong scattering, the value of $f_{\rm pinhole}$ decreases with increasing $M_{\bullet}$ and drops steeply for $ M_{\bullet} \sim 10^8 M_{\sun}$, as expected for cusp galaxies. Interestingly, the influence of strong scattering modifies the pinhole fraction in two opposite directions, with the net effect depending on $\gamma_{\rm bh}$. The first effect is already explained in the preceding paragraph: at fixed $\tau$, loss cone flux is more greatly reduced (by strong scatterings) when $\epsilon$ is low than when $\epsilon$ is high.  However, a countervailing force is visible in Fig. \ref{fig:FK}, where we see the mild but relevant time evolution of the suppression due to strong scatterings.  The suppression factor $\mathcal{F}(\epsilon)/\mathcal{F}_{\rm w}(\epsilon)$ becomes smaller and smaller as $\tau$ grows, so when computing an energy-integrated $\dot{N}_{\rm TDE}$ rate, the bins of large $\epsilon$ (with larger $\tau$) will be further along the time evolution of their suppression curves.  In Fig. \ref{fig:pinhole}, we see that for large $\gamma_{\rm bh}$, the first effect wins out and the pinhole fraction decreases due to ejections.  But for $\gamma_{\rm bh}\lesssim 9/4$, the second effect wins out and the pinhole fraction sees little net change from strong scatterings (the net change that does exist is a very mild increase in $f_{\rm pinhole}$).
 
\subsection{E+A preference}
Observations have revealed that E+A and other post-starburst galaxies are significantly over-represented among TDE hosts \citep{Arcavi+14, French+16,French+17, Law, Graur, French20, Hammerstein21}. 
A few solutions have been proposed to explain this preference: stellar overdensities \citep{Stone+18}, radial anisotropies  \citep{Stone+18}, SMBH binaries (e.g., 
\citealt{Arcavi+14}) and accounting for a complete stellar mass function \citep{Bortolas}. 
\\
In the framework of our revised loss cone theory, we have shown that steep slopes induce a strong reduction of TDE rates. Motivated by our findings, here we study the impact of stellar overdensities with strong scattering. 
\\
E+A galaxies are post-starburst galaxies and hence, in this section, we consider only stars and no sBHs. If we assume that most of the central stars formed impulsively in a starburst, these ultra-steep stellar central density cusps will relax over time towards a steady state configuration: the Bahcall–Wolf cusp, with $\gamma_{\star} = 7/4$ \citep{BW}.
An ultra-steep stellar density profile will result in much shorter relaxation times, especially at small energies, and hence the relaxation to a BW cusp is quicker for steeper star profiles. Following \citet{Stone+18}, we define the Bahcall-Wolf radius $r_{\rm BW}$ as the radius where the post-starburst age $t$ equals the local energy relaxation time: 
\begin{equation}
r_{\rm BW} =\left( \frac{M_{\bullet}^{3/2} \langle m_{\star} \rangle }{G^{1/2} \langle m^2_{\star}\rangle  \rho_{inf} r_{inf}^{\gamma} \: t \: \ln \Lambda}  \right) ^{\frac{1}{3/2- \gamma}} 
\end{equation}

\begin{figure}
	\centerline{\includegraphics[width=90mm]{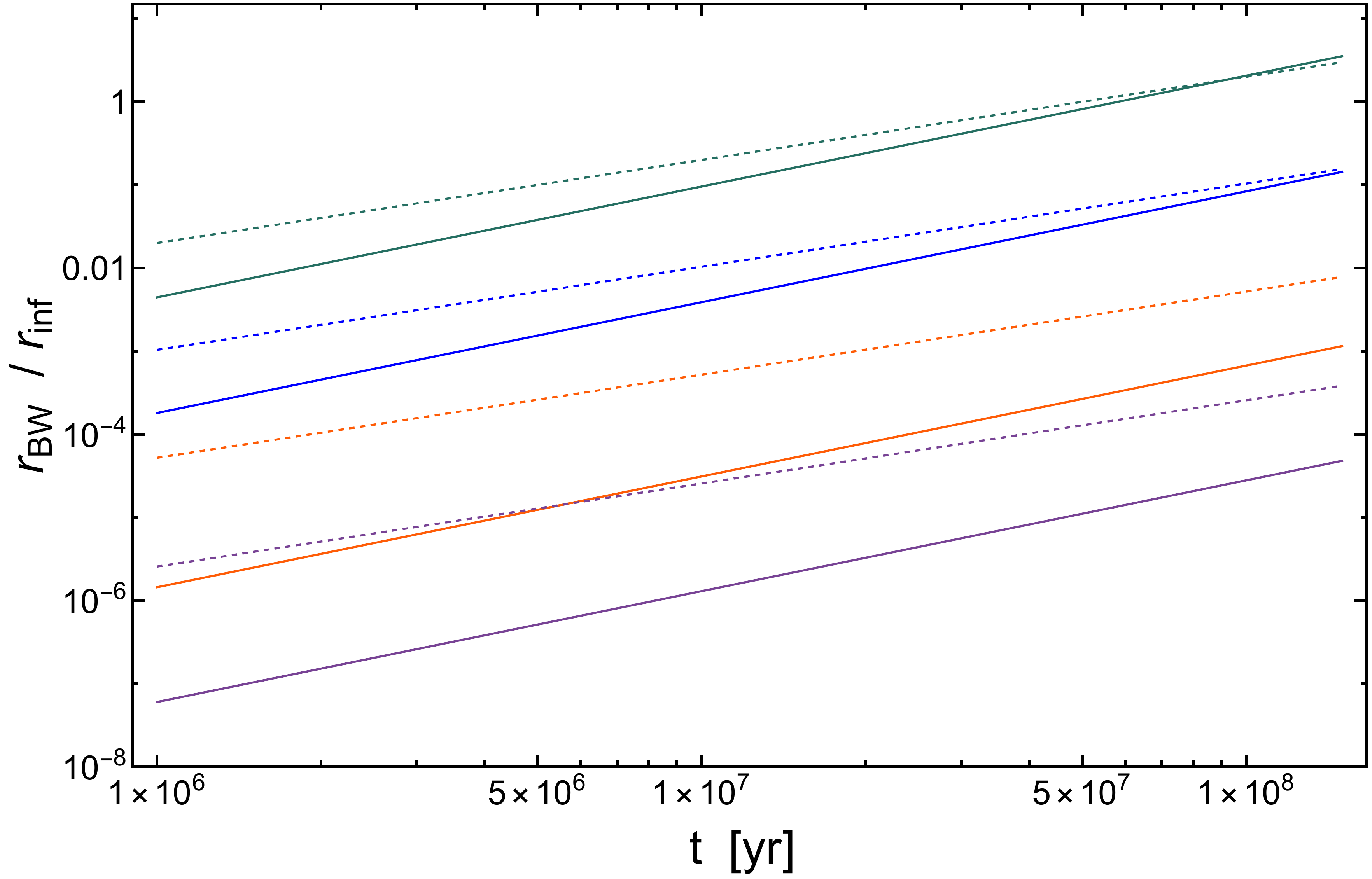}}
    \caption{Time evolution of the Bahcall-Wolf radius divided by the radius of influence for different MBH masses. Green, blue, orange, and purple lines correspond to $M_{\bullet}= 10^5 M_{\sun}$, $M_{\bullet}= 10^6 M_{\sun}$, $M_{\bullet}= 10^7 M_{\sun}$, and $M_{\bullet}= 10^8 M_{\sun}$, respectively.  The dashed lines correspond to $\gamma_{\star}= 5/2 $ while full lines are for  $\gamma_{\star}= 9/4 $. }
   \label{fig:rbw}
\end{figure}

In Fig. \ref{fig:rbw}, we plot the Bahcall-Wolf radius $r_{\rm BW}$ divided by the influence radius for different MBH masses. 
As $r_{\rm BW}$ expands, energy relaxation erodes the population of the most tightly bound stars, decreasing their density relative to the initial conditions.  The density decrease at high $\epsilon$ reduces the efficiency of loss cone shielding, as it is the most tightly bound stars which are responsible for the strongest shielding effects. From Fig. \ref{fig:rbw}, we see that this cusp erosion is fastest for small MBHs and slowest for large MBHs.  As we do not self-consistently model this cusp erosion in our calculations, we only compute TDE rate reductions at times where this effect would be minor. 

\begin{figure}
	\centerline{\includegraphics[width=90mm]{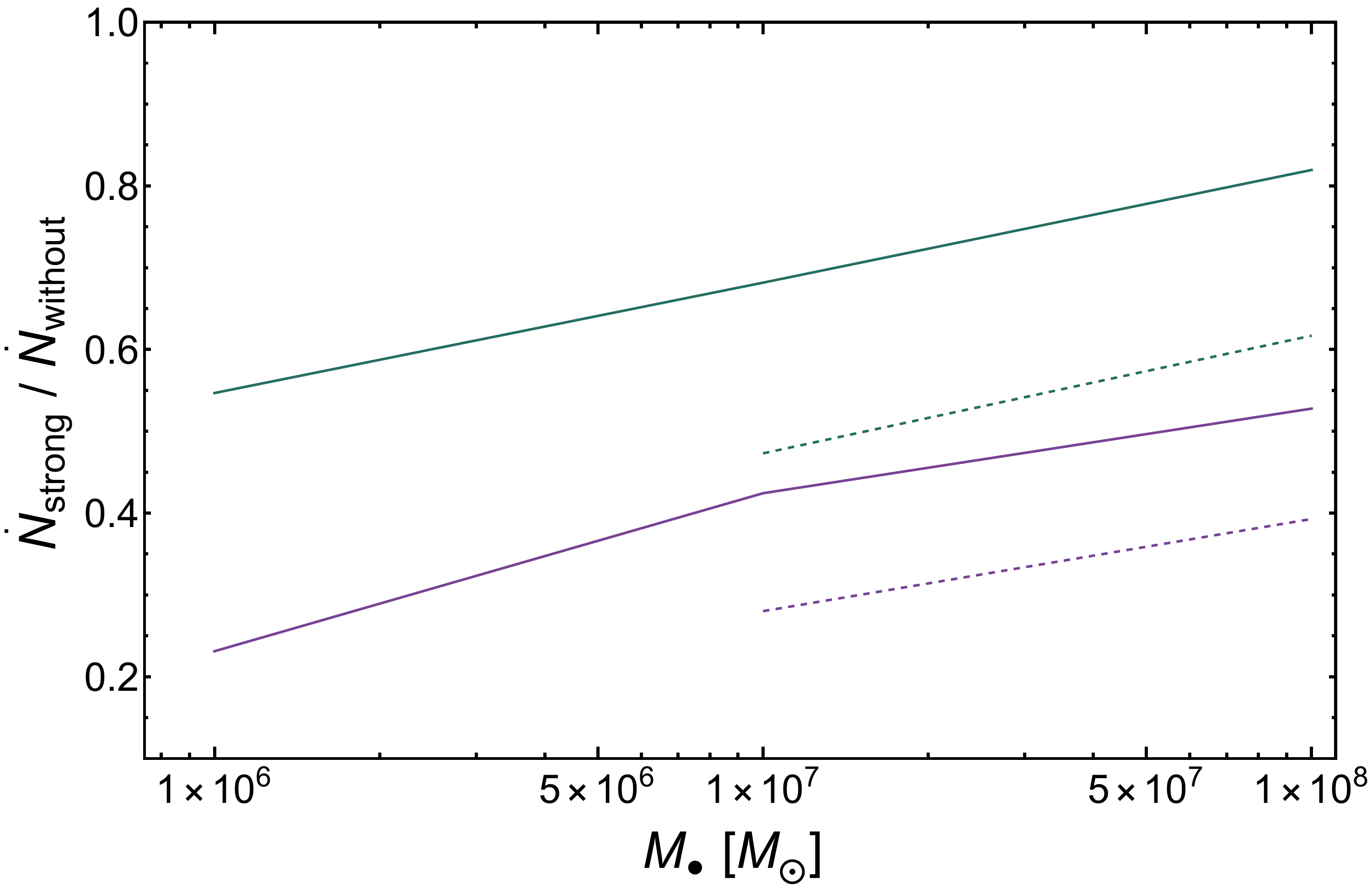}}
    \caption{Rates of TDEs with strong scattering divided by the rates of TDEs without strong scattering, shown as a function of the MBH mass. Green lines are for $\gamma_{\star}= 9/4 $, purple lines for $\gamma_{\star}= 5/2 $. Full lines are at $t= 10$ Myr (15 Myr for  $\gamma_{\star}=9/4 )$, dashed lines are at $t= $100 Myr (150 Myr for  $\gamma_{\star}=9/4 )$.}
   \label{fig:ea}
\end{figure}

\rev{In Fig. \ref{fig:ea}, we present TDE rates that consider the influence of strong scattering, normalized by TDE rates that do not account for strong scattering. These rates are depicted at two different times where the cusp erosion can be disregarded: $\frac{r_{BW}}{r_{inf}} \leq 10^{-3}$,} \footnote{\rev{This criterion derives from the fact that energies such as 
$r \leq 10^{-3} r_{inf}$ contribute about 1\% of the TDE rate.}}
\rev{ (see Fig. \ref{fig:rbw}). We find that, for the steep stellar profile: $\gamma \geq 9/4$, accounting for strong scattering can reduce TDE rates by up to an order of magnitude. Additionally, it is worth noting that the effect is more pronounced for the dashed lines corresponding to times $t=100$ Myr and  $t=150$  Myr (the cusp erosion is still negligible for the higher MBH masses). Our results suggest that ultra-steep stellar profiles can effectively self-shield, which makes them a less plausible explanation for the preference of TDEs for E+A galaxies.}

\section{Discussion \& Conclusions}
\label{sec:conclusions}

In this paper, we have proposed a revised loss cone theory that takes into account both classical weak scattering and close encounters: strong scattering, collisions, tidal captures and $\mu$-TDEs. These close encounters can occur at non-negligible rates inside the radius of influence, 
where stellar densities can be higher than anywhere else in the universe.  Below we summarize our key findings: 
\\
\begin{enumerate}
    \item For stars with semi-major axis $a \sim r_{\rm inf} $, i.e. the primary sources of TDEs, strong scattering rates are significantly larger than other close encounter rates. For very high orbital energies, collisions become the dominant close encounter. However, those energies have a very small contribution to TDE rates.  
  \item We derive time-dependent analytical solutions of the Fokker-Planck equation with strong scattering for 
  a broad range of physically motivated slopes. This family of analytic solutions, which we can write in closed form for all relevant integer, half-integer, and quarter-integer $\gamma$ values, is a generalization of the standard \cite{CohnKulsrud78} $f(j)$ distribution function in the presence of strong scatterings. 
  \item We find that for black hole slopes  $\gamma_{\rm bh} \geq 9/4 $, the loss cone will be effectively shielded by strong scattering. TDE rates are reduced by a factor $\sim 2 $ for $\gamma_{\rm bh} = 9/4 $ and by a factor $\geq 10 $ for $ \gamma_{\rm bh} = 5/2$.  
  \item Strong scattering is more effective at removing stars with $a \sim r_{\rm inf}$ compared to stars with $a \ll r_{\rm inf}$. Thus, in addition to shielding the loss cone, strong scattering can also reduce the fraction of TDEs in the pinhole regime, and therefore the fraction of deeply plunging TDEs for $\gamma_{\rm bh} = 5/2$. 
  \item 
  \rev{Stellar overdensities have been proposed as one possible explanation for the observed preference of tidal disruption events (TDEs) in E+A galaxies. However, our revised loss cone theory challenges this idea. In classical loss cone theory, ultra-steep stellar slopes increase the diffusion coefficients and, thus, the TDE rates. However, we have shown that they also increase the rates of strong scattering, which can reduce TDE rates. As a result, the net enhancement of TDEs induced by stellar overdensities is milder than previously assumed. This finding makes other possible explanations, such as radial anisotropies (e.g., \citealt{Stone+18}) or a more complete mass function (e.g., \citealt{Bortolas}), more likely to explain the observed preference of TDEs for E+A galaxies.}
\end{enumerate}

In this paper, we have focused on the most dramatic qualitative difference between weak and strong scatterings: ejection.  However, stars also have mildler but still strong encounters that can non-diffusively move them to new -- but still bound -- energies.  
This effect will be the subject of a future work. The shielding effect we have studied depends on black hole slopes which have not been constrained observationally. $N$-body simulations have shown that sBHs 
can settle to a slope of $\gamma_{\rm bh} \geq 2$ \citep{Amaro,Preto}, in agreement with some predictions of strong mass segregation \citep{AlexanderHopman09}. \rev{It can also be noted that, for an  adiabatically growing MBH, the stellar slope may assume steeper values  of $\gamma \gtrsim 2$ \citep{Young}.} Moreover, \cite{Generozov+18} showed that, in the presence of a source term that is continuously depositing new stars or sBHs at small radii, black holes will settle into a density profile with $ \gamma_{\rm bh} = 5/2$. Motivated by these studies, we considered different black hole slopes with $\gamma_{\rm bh} \leq 2.5 $.  
\\
If steep profiles for black holes are common in galactic nuclei, this may explain persistent discrepancies between the high predicted rates of TDEs \citep{WangMerritt04} and the (relatively) low observed rates of TDEs in low-mass galactic nuclei.  While the steep TDE luminosity function \citep{vanVelzen18} helps to resolve this rate discrepancy, it may be unable to resolve the entire breadth of the discrepancy if the occupation fraction of small MBHs ($\sim 10^5 M_\odot$) is high in dwarf galactic nuclei.  Conversely, loss cone shielding may pose a challenge for the ``overdensity'' solution \citep{StoneMetzger16, StonevanVelzen16, Stone+18} to the observed post-starburst preference \citep{Arcavi+14, French+16, French+17,Hammerstein21} among the host galaxies of TDEs.  While an ultra-steep stellar distribution can boost TDE rates up to a point, our work here shows that the gains from increasing $\gamma_\star$ beyond $9/4$ may be self-limiting, and that TDE rates may even drop as $\gamma_\star$ increases to very large values.
\\
It is worth mentioning that we considered a monochromatic distribution of stars throughout the paper. However, we did also compute the impact of a Kroupa IMF and found that it increased all TDE rates with and without strong scattering by a common factor of 1.6. 
It is also possible that shielding effects could be important for the production of hypervelocity stars in binary breakups \citep{Hills88}, as these are generally sourced from binaries with galactocentric semimajor axes $a \gg r_{\rm infl}$ \citep{Perets+07}.
\\
To conclude, we emphasize that we considered the 1D Fokker-Planck equation in angular momentum, neglecting energy evolution of the stellar (and sBH) distribution function. This may be important for steep density profiles that are not quasi-steady states. This, together with the effect of strong scattering that can non-diffusively move stars to higher energies will be the subject of a future work.

\section*{Acknowledgements}
The authors would like to thank Cole Miller, Eliot Quataert and Luca Broggi for helpful discussions. We also warmly thank the  anonymous referee for their very useful comments and suggestions. OT gratefully acknowledges the support of the Einstein-Kaye scholarship and of the Israel Ministry of Science and Technology. This work was partially funded by the Israel Science Foundation (Individual Research Grant 2565/19) and the United States-Israel Binational Science Foundation (Grant No. 2019772). 

\section*{DATA AVAILABILITY STATEMENT}
The data underlying this article will be shared on reasonable request to the corresponding author.

\bibliographystyle{mnras}
\bibliography{lc_refs} 

\appendix
\onecolumn

\section{Orbit-averaged ejection rate}
\label{app:analytics}
Throughout this paper, we performed orbit averages of various interaction rates $X$ in the following way:
\begin{equation}
    \langle X \rangle = \frac{2}{P} \int_0^\pi \dot{\nu}^{-1} X {\rm d}\nu.
\end{equation}
In the special case of a Kepler potential, this simplifies to
\begin{equation}
    \langle X \rangle = \frac{2(1-e^2)^{3/2}}{\pi} \int_0^\pi \frac{X{\rm d}\nu}{(1+e\cos \nu)^2} . \label{eq:orbitAvg}
\end{equation}
We found that the orbit averaged ejection rate has an analytical form for $\gamma_\star$ values which are integers or half-integers. Hereafter are the analytic forms we found for interesting values of $\gamma$. 
For an equal mass scatterer and $ \gamma_\star = 3/2$, we obtained: 
 \begin{equation}
 \langle \dot{N}_{\rm ej} \rangle= \frac{2^{1/2} \pi \rho_{\rm infl} m G^{1/2} r_{\rm infl}^{3/2} (4 - 3 (1 - e^2)^{1/2})}{ (\gamma_\star+ 1)M ^{3/2}  (1-e^2)^{1/2}} .
\end{equation}
\\
For an equal mass scatterer, with $\gamma_\star= 5/2$, we obtained: 
 \begin{equation}
 \langle \dot{N}_{\rm ej} \rangle= \frac{2^{-1/2}\pi \rho_0 m G^{1/2} r_0^{5/2} (-4+ 12 e^2 + 5 (1-e^2)^{3/2})}{(\gamma_\star + 1) M ^{3/2} a (1-e^2)^{3/2}} .
\end{equation}

\section{Local diffusion coefficient }
\label{app:diff}
The local diffusion coefficient we evaluate in Eq. \ref{eq:diffAvg} is given by \citep{MagorrianTremaine99, WangMerritt04}: 
\begin{equation}
\lim _{R \rightarrow 0} \frac{\left\langle(\Delta R)^{2}\right\rangle}{2 R}=\frac{32 \pi^{2} r^{2} G^{2}\left\langle m_{\star}^{2}\right\rangle \ln \Lambda}{3 J_{c}^{2}(\epsilon)}\left(3 I_{1 / 2}(\epsilon)-I_{3 / 2}(\epsilon)+2 I_{0}(\epsilon)\right)
\end{equation}
with: 
\begin{equation}
I_{0}(\epsilon) \equiv \int_{0}^{\epsilon} f\left(\epsilon^{\prime}\right) \mathrm{d} \epsilon
\end{equation}
\\
and 
\begin{equation}
I_{n / 2}(\epsilon) \equiv  {[2(\psi(r)-\epsilon)]^{-n / 2} } 
\int_{\epsilon}^{\psi(r)}\left[2\left(\psi(r)-\epsilon^{\prime}\right)\right]^{n / 2} f\left(\epsilon^{\prime}\right) \mathrm{d} \epsilon^{\prime}.
\end{equation}

 \end{document}